\pdfoutput=1
\documentclass[12pt]{article}
\usepackage{jheppub}
\pdfoutput=1
\usepackage{epstopdf}
\usepackage{mathrsfs}
\usepackage{psfrag}
\usepackage{color}
\usepackage{hyperref}
\usepackage{empheq}
\usepackage{slashed}
\usepackage{simplewick}
\usepackage{cancel}
\usepackage[utf8]{inputenc}
\usepackage[table]{xcolor}
\usepackage{amsmath,bbm,array,amsfonts,graphicx,wrapfig,arydshln,lscape,float,multirow,longtable,rotating,makecell}
\usepackage{url}
\usepackage{caption}
\usepackage{subcaption}
\usepackage[backgroundcolor=black!5, linecolor=green!60]{todonotes}
\hypersetup{pdftitle={},pdfcreator={},linkcolor=[rgb]{0.15,0.35,0.75},colorlinks=true,citecolor=[rgb]{0.675,0,0.2},urlcolor=[rgb]{0.15,0.35,0.65}}
\let\olditemize\itemize\renewcommand{\itemize}{\vspace{-2pt}\olditemize\setlength{\itemsep}{1pt}\setlength{\parskip}{0pt}\setlength{\parsep}{-0pt}}
\let\oldenumerate\enumerate\renewcommand{\enumerate}{\vspace{-4pt}\oldenumerate\setlength{\itemsep}{1pt}\setlength{\parskip}{0pt}\setlength{\parsep}{0pt}}

\def\beq{\begin{equation}}
\def\eeq{\end{equation}}
\def\bsp#1\esp{\begin{split}#1\end{split}}
\newcommand{\be}{\begin{equation}}
\newcommand{\ee}{\end{equation}}
\newcommand{\bea}{\begin{eqnarray}}
\newcommand{\eea}{\end{eqnarray}}
\newcommand{\eqn}[1]{eq.~\eqref{#1}}

\def\eqn#1{eq.~(\ref{#1})}
\def\eqns#1#2{eqs.~(\ref{#1}) and~(\ref{#2})}

\def\spa#1.#2{\left\langle#1\,#2\right\rangle}
\def\spb#1.#2{\left[#1\,#2\right]}
\newcommand{\e}{\epsilon}
\def\nn{\nonumber}
\def\Li{{\rm Li}}

\def\cO{{\mathcal O}}
\def\cL{{\mathcal L}}

\def\lr{\leftrightarrow}
\renewcommand{\phi}{\varphi}
\renewcommand{\bar}{\overline}

\newcommand{\ab}[1]{\langle #1 \rangle}
\newcommand{\sqb}[1]{[ #1 ]}

\newcommand{\sab}[1]{s_{#1}}

\newcommand{\lam}[1]{\lambda_{#1} }
\newcommand{\lamt}[1]{\widetilde \lambda_{#1} }

\def\lra{\leftrightarrow}

\def\N{\mathcal{N}}

\def\tr{\text{tr}}

\def\ksl{\not{\hbox{\kern-2.3pt $k$}}}

\def\e{\epsilon}

\def\Ord{{\cal O}}

\def\bom#1{{\mbox{\boldmath $#1$}}}

\def\lr{\leftrightarrow}

\def\Li{\mathop{\rm Li}\nolimits}

\def\eqn#1{eq.~(\ref{#1})}
\def\eqns#1#2{eqs.~(\ref{#1}) and~(\ref{#2})}

\def\spa#1.#2{\left\langle#1\,#2\right\rangle}
\def\spb#1.#2{\left[#1\,#2\right]}
\def\lor#1.#2{\left(#1\,#2\right)}
\def\sand#1.#2.#3{%
\left\langle\smash{#1}{\vphantom1}^{-}\right|{#2}%
\left|\smash{#3}{\vphantom1}^{-}\right\rangle}

\def\eg{{\it e.g.~}}
\def\ie{{\it i.e.~}}

\newcommand{\mcdot}{\!\cdot}

\newcommand{\zb}{\bar{z}}
\def\Vol#1#2{\left[\frac{ \mu^2 (-s_{#1 #2})}{(-s_{#1 q})(- s_{q #2} )}\right]}
\def\im{\mathrm{i}}

\newcommand{\NeqFour}{\mathcal{N}\!=\!4\text{ SYM}}

\definecolor{airforceblue}{rgb}{0.36, 0.54, 0.66}
\definecolor{bananayellow}{rgb}{1.0, 0.88, 0.21}
\definecolor{bittersweet}{rgb}{1.0, 0.44, 0.37}
\definecolor{blue(ncs)}{rgb}{0.0, 0.53, 0.74}
\definecolor{bole}{rgb}{0.47, 0.27, 0.23}
\definecolor{brass}{rgb}{0.71, 0.65, 0.26}
\definecolor{bronze}{rgb}{0.8, 0.5, 0.2}
\definecolor{brgreen}{rgb}{0.0, 0.26, 0.15}
\definecolor{burgundy}{rgb}{0.5, 0.0, 0.13}
\definecolor{cherry}{rgb}{1.0, 0.72, 0.77}
\definecolor{cocao}{rgb}{0.82, 0.41, 0.12}
\definecolor{citrine}{rgb}{0.99, 0.82, 0.07}

\makeatletter\DeclareRobustCommand*{\bfseries}{\not@math@alphabet\bfseries\mathbf\fontseries\bfdefault\selectfont\boldmath}\makeatother

\title{Soft gluon emission at two loops in full color}
\author[1]{Lance~J.~Dixon,}
\affiliation[1]{SLAC National Accelerator Laboratory, Stanford University, Stanford, CA 94039, USA}
\emailAdd{lance@slac.stanford.edu}

\author[1]{Enrico Herrmann,}
\emailAdd{eh10@stanford.edu}

\author[2]{Kai~Yan,}
\affiliation[2]{Max-Planck-Institut f\"ur Physik, Werner-Heisenberg-Institut, D-80805 M\"unchen, Germany}
\emailAdd{kyan@mpp.mpg.de}

\author[3]{and Hua~Xing~Zhu}
\affiliation[3]{Zhejiang Institute of Modern Physics, Department of Physics, Zhejiang University, \\
Hangzhou, 310027, China}
\emailAdd{zhuhx@zju.edu.cn}

\preprint{SLAC--PUB--17480,\ \ MPP-2019-255}

\abstract{
  The soft emission factor is a central ingredient in the factorization of
  generic $n$-particle gauge theory amplitudes with one soft gluon
  in the external state.  We present the complete two-loop soft factor,
  capturing the leading power behavior in the soft-gluon momentum.
  At two loops, the color structure and the kinematic dependence of
  the soft factor become nontrivial as the soft gluon can couple to three 
  hard partons for the first time (tripole terms).  The nontrivial kinematic 
  dependence of the tripole terms is of uniform, maximal transcendental
  weight, and can be expressed (in a ``Euclidean'' region) in
  terms of single-valued harmonic polylogarithms.  Our results are
  consistent with the behavior of the recently computed symbol
  of the two-loop five-particle amplitude in $\mathcal{N}=4$
  super-Yang-Mills theory. In the limit where the outgoing soft gluon
  is also collinear with an incoming hard parton,
  potentially dangerous factorization-violating terms can arise,
  but they cancel after summing over colors. }

\begin{document}
\maketitle

%================================================
\vspace{-20pt}
%\newpage
\section{Introduction}
\label{sec:intro}\vspace{-8pt}
%================================================
%
Scattering amplitudes are some of the most intensively studied
quantities in gauge theory. On the one hand, they are essential
building blocks for precision QCD predictions at high energy
colliders, such as the Large Hadron Collider. On the other hand, their
analytic structure often reveals hidden properties of gauge theory
which are otherwise hard to
probe~\cite{Drummond:2006rz,Alday:2007hr,Drummond:2008vq}. 
Due to their fundamental importance, there have been significant
efforts to understand and exploit the analytic properties
of scattering amplitudes.

Multi-loop scattering amplitudes are complicated functions of the
momenta and helicities of the external particles.
In certain kinematical regimes, gauge theory
amplitudes become simpler, by factorizing into products
of full lower-point amplitudes~\cite{Mangano:1990by},
or into lower-point amplitudes multiplied by a universal emission factor.
Examples of the latter type include collinear
factorization~\cite{Mangano:1990by,Bern:1994zx,Dixon:1996wi},
where two or multiple partons become collinear, or soft factorization,
where one or more partons become
soft~\cite{Bassetto:1984ik,Berends:1988zn,Bern:1999ry}.
Factorization is of great theoretical interest for several reasons:
\begin{itemize}
\item[a)] The computation of perturbative cross sections in gauge theory at
higher order in the coupling constant often involves phase space
integrals over emission factors arising in the factorization
limit~\cite{Giele:1991vf,Giele:1993dj,Campbell:1997hg,Catani:1999ss}.
\item[b)] Certain observables exhibit large logarithms at fixed order,
  invalidating a simple perturbative calculation. Accurately describing
  these observables requires resumming the large logarithms,
  which relies on
  factorization~\cite{Collins:1989gx,Sterman:1995fz,Sterman:2004pd}.  
\item[c)] In some cases, complete scattering 
amplitudes can be ``bootstrapped'' from their asymptotic factorization
behavior~\cite{Bern:1993qk,Bern:1997sc,Dixon:2013eka,Caron-Huot:2019vjl}.
\end{itemize}
In this work, we focus on the leading power behavior of gauge theory
amplitudes when a single soft gluon is emitted.
At tree level, this corresponds to the 
simple and well-known eikonal approximation~\cite{Weinberg:1965nx}. 
At one loop, due to infrared divergences, the soft factor receives
perturbative
corrections~\cite{Bern:1995ix,Bern:1998sc,Bern:1999ry,Catani:2000pi},
which can couple at most two hard partons at this order.
At two loops, the soft factor has been known for some time for
collision processes with two hard colored external
states~\cite{Li:2013lsa,Duhr:2013msa}; the same
function also describes soft emission in the (planar) limit of
a large number of colors $N_c$ for gauge group SU$(N_c)$.
However, starting at two loops, it is possible to couple up to three
hard particles in the single soft emission process, in a non-planar
fashion that contributes at subleading order in $N_c$.
One of the main novel results in this paper is the computation of this
additional ``tripole'' contribution to the soft factor. 

Because the tripole contribution contains only soft gluons, it is
the same in any gauge theory at the lowest order of its appearance, namely two loops in perturbation theory. 
In particular, it is the same in QCD
as in $\mathcal{N}=4$ super-Yang-Mills theory (SYM).  Therefore
the ``uniform transcendentality'' properties of $\mathcal{N}=4$ SYM
amplitudes~\cite{Dixon:2011pw,BDS,ArkaniHamed:2012nw,LipatovTranscendentality}
imply similar properties for the universal tripole contributions
computed here.  (The non-tripole contributions
depend on the matter content of the theory and do not have uniform
transcendentality at two loops~\cite{Li:2013lsa,Duhr:2013msa}.)
Also, we can employ our results to
verify the soft-gluon limit of the full-color two-loop
five-particle amplitude in $\mathcal{N}=4$ SYM, which was
computed recently at the level of the
symbol~\cite{Abreu:2018aqd,Chicherin:2018yne}.

By considering soft-gluon emission in a direction collinear to one of
the external hard particles, we can access the soft limit of the
collinear splitting amplitudes. In particular, we obtain results in
the spacelike splitting regime which are so far, apart from the
singular terms, unknown.  Given the one- and two-loop soft factor in
the spacelike collinear limit, we
investigate
the breaking of strict
collinear factorization at next-to-next-to-leading order parton splitting.
Such factorization breaking has been observed previously
for splitting amplitudes at the level of infrared-divergent
terms, which were argued to cancel in color-summed, 
squared QCD matrix elements~\cite{Catani:2011st,Forshaw:2012bi}
at this order (although perhaps not at the next
order~\cite{Forshaw:2006fk,Forshaw:2008cq,Forshaw:2012bi}).
Here we extend these results to the {\it infrared-finite}
terms in dimensionally-regularized amplitudes
and squared splitting probabilities.  We find that potential
factorization-violating terms
cancel at the level of the cross section.

The remainder of this work is structured as follows: 
In section \ref{sec:soft_factorization} we briefly summarize the 
relevant features of soft-gluon factorization, and define our conventions
while reviewing the known tree-level and one-loop results.
Section \ref{sec:soft_factorization_2L} constitutes the central part
of this paper. First, we define the decomposition of the two-loop soft factor 
into dipole terms (known in the literature) and tripole terms.
After summarizing the relevant tripole kinematics,
the new pieces are given in one representation
in section \ref{subsec:tripole}, in a ``Euclidean" kinematic configuration.
We provide an alternate representation for the dipole terms in
section \ref{subsec:tripolealt}, eqs.~(\ref{tripolesumALT})
and (\ref{eq:tripolesum}), still in the Euclidean region.
In section \ref{subsec:analytic_cont}, we define the analytic continuation
to all other kinematic configurations. This continuation will play a role in
our investigation of strict collinear factorization violation at
cross-section level, as we take the soft factor into a (spacelike) collinear
limit in section \ref{subsec:soft_collinear}.
We also mention the relevance of the soft factor for non-planar
two-loop five-particle amplitudes in section \ref{subsec:soft_lim_amps},
and finally we conclude in section \ref{sec:conclusions}.
We provide an ancillary file containing computer-readable
versions of some of the lengthier expressions in this paper.

%================================================
\vspace{-0pt}
\section{Soft factorization}
\label{sec:soft_factorization}\vspace{-8pt}
%================================================
%
Consider a scattering process with $n+1$ all outgoing\footnote{Processes with incoming colored partons can be
obtained from the all-outgoing case by crossing.} colored partons and any
number of color-neutral particles, 
\begin{align}
0 \to p_1 + p_2 +\dots + p_n + q + \textrm{color neutral particles}\,,
\end{align}
where $q$ denotes the momentum of the soft gluon. In the
limit where $q$ becomes soft, \ie $q\cdot p_i$ is parametrically
smaller than $p_i \cdot p_j$ for all $1\leq i, j \leq n$, the
(bare, unrenormalized)~amplitude for this process factorizes into
the product of a singular soft
factor and an amplitude omitting the soft gluon\footnote{We include the definition 
of the soft current $\bom{J}_{\!\mu}(q)$ for convenient comparison 
with \eg ref.~\cite{Duhr:2013msa}.}:
\begin{align}
  \label{eq:1}
  \big| \mathcal{M}_{n+1}(\{p_m\}, q) \big\rangle 
  \simeq 
  \epsilon^\mu_{\pm}(q) \bom{J}_{\!\mu}(q)  \big|\mathcal{M}_{n}(\{p_m\}) \big\rangle
  \equiv
  \bom{S}^\pm(\{\beta_m\},q) \big|\mathcal{M}_{n}(\{p_m\}) \big\rangle \, .
\end{align}
For an SU$(N_c)$ gauge theory such as QCD, the amplitude can be
conveniently represented as a vector in color space~\cite{Catani:1996vz}. 
The soft factor, $\bom{S}^\pm(\{\beta_m\},q)$, and likewise the soft current $ \bom{J}_{\!\mu}(q)$, are 
color operators passing from an $n$-particle color space to an
$(n{+}1)$-particle color space by adding one soft gluon~\cite{Catani:2000pi}.
In eq.~(\ref{eq:1}), the `$\simeq$' sign signals that the equality 
holds for the leading terms in the soft limit as the gluon's energy vanishes, 
$q^0 = |\vec{q}| \to 0$, up to power corrections in $q^0$.  
Also, `$\pm$' in $\e_{\pm}(q)$ denotes the polarization (helicity) of the soft gluon. 
The (leading) soft factorization is universal in the sense that the
soft factor is independent of the helicities and flavor of the hard partons
in the process; it only depends on the color charge and angular direction
of the hard partons, $\beta^\mu_m = p^\mu_m/p^0_m$.  (In this work we consider
massless partons only, and just the leading powers in $q^0$.)
As we shall see later, the latter fact is particularly important
and is sometimes dubbed ``rescaling
invariance''~\cite{Kidonakis:1998nf,Aybat:2006mz,Dixon:2008gr,%
  Gardi:2009qi,Becher:2009cu,Almelid:2015jia},
or ``reparameterization invariance''~\cite{Manohar:2002fd}.

The $n$-point, all-gluon amplitudes can be expanded perturbatively in the
strong coupling constant $g_s$,
\begin{align}
  \big|\mathcal{M}_{n}(\{p_m\}) \big\rangle & =
  \sum^\infty_{L=0} g^{n-2}_s \, \bar{a}^L \,
  \big|\mathcal{M}^{(L)}_{n}(\{p_m\}) \big\rangle \,, 
  \label{eq:2}
\end{align}
where we have introduced a rescaled coupling,
\beq
  \bar{a}\ \equiv \frac{g_s^2}{(4\pi)^{2-\e}} \, e^{-\epsilon \gamma_E}
    \ =\ \frac{\alpha_s}{4\pi} \, \frac{e^{-\epsilon \gamma_E}}{(4\pi)^{-\e}} \,,
\label{eq:abardef}
\eeq
where $\alpha_s = g^2_s/(4\pi)$,
and $\gamma_E$ is the Euler-Mascheroni constant.
Perturbatively expanding the factorization formula (\ref{eq:1}) subsequently
defines the perturbative expansion of the soft factor,
\begin{align}
 g^{(n{+}1){-}2}_s \, \bar{a}^L \, \big| \mathcal{M}^{(L)}_{n{+}1}(\!\{p_m\},q)\big\rangle  
 & \simeq
 \sum^L_{\ell=0} g_s \bar{a}^{\ell} \bom{S}^{\pm,(\ell)}(\!\{\beta_m\},q)  
 			  \! \times  \!
g^{n{-}2}_s \, \bar{a}^{L{-}\ell} \,
  \big| \mathcal{M}^{(L{-}\ell)}_{n}(\!\{p_m\}\!)\big\rangle \,, 
  \\
\bom{S}^\pm(\!\{\beta_m\},q)  & \equiv
  \sum^{\infty}_{L=0} g_s \, \bar{a}^{L} \, \bom{S}^{\pm,(L)}(\!\{\beta_m\},q) \,.
\end{align}
It has been shown diagrammatically that eq.~(\ref{eq:1}) holds
at one loop in ref.~\cite{Catani:2000pi},
and later to all loops in refs.~\cite{Feige:2014wja,Larkoski:2014bxa}.
In particular, the soft factor can be calculated to all orders
as the matrix element of time-ordered Wilson line operators with
a single gluon in the external state:
\begin{align}
  \label{eq:WilsonLineRep}
  \bom{S}^\pm(\{\beta_m\},q)
  = \langle q | \int \! d^4 x \, e^{\im x \cdot q} \, \mathrm{T} \Bigl\{  \prod_{k=1}^n Y_k(x) \Bigr\} | 0 \rangle \,,
\end{align}
where $Y_k(x)$ is a semi-infinite Wilson line that acts as a
lightlike color source for one of the $n$ hard partons in the external
state.

For example, for an outgoing parton with momentum $p_k$, its corresponding 
Wilson line starts from the scattering origin, $x$, and extends to
null infinity along the direction of the parton velocity $\beta_k$:
\begin{align}
  \label{eq:26}
  Y_k(x) = P\exp\Big(\im \frac{g_s}{\sqrt{2}}\! \int^\infty_0 \! ds\, \beta_k \cdot A^a(x + s\beta_k)\, \bom{T}^a_k \Big) \,.
\end{align}
Here, $P$ stands for path ordering, and $\bom{T}^a_k$ is the color-charge
operator in the color space formalism~\cite{Catani:1996vz}. For an outgoing quark~(incoming
anti-quark), $\big(T^a_k\big)_{\alpha\beta} = (t^a)_{\alpha\beta}$, for an
outgoing anti-quark~(incoming quark), $\big(T^a_k\big)_{\alpha\beta} =
(-t^a)_{\beta\alpha}$, and for a gluon, $\big(T^a_k\big)_{\alpha\beta} = f^{\alpha a \beta}$, 
where $t^a$ are the \emph{non}-standard Gell-Mann matrices normalized according to 
\begin{align}
  \tr[t^a t^b] = \delta^{ab}\,, \text{ satisfying the color-algebra } [t^a, t^b]
  = f^{abc} t^c\,.
\end{align}
Note that the structure constants $f^{abc}$ used in this paper are
$\im\sqrt{2}$ times larger than the conventional ones, and are often denoted by
$\tilde{f}^{abc}$, but here we will drop the `$\sim$'.
Our normalization of the generators $T^a_k$ also accounts
for the $1/\sqrt{2}$ in \eqn{eq:26}.

At tree level, the soft current is simply the well-known eikonal factor
\begin{align}
  \bom{J}^{a,(0)\,\mu}(q)
  = \frac{1}{\sqrt{2}}\sum^n_{i=1} \bom{T}^a_i \frac{p^\mu_i}{p_i \cdot q} 
  = \frac{1}{\sqrt{2}} \frac{1}{2n} \sum^n_{i\neq j =1 } 
  \big(\bom{T}^a_i - \bom{T}^a_j \big)
  \left( \frac{p^\mu_i}{p_i \cdot q}  - \frac{p^\mu_j}{p_j \cdot q}  \right)\,,
\end{align}
which can be shown by using color conservation, $\sum^n_{i=1} \bom{T}^a_i =0$.
The factor of $1/(2n)$ removes overcounting.  
 
In terms of spinor helicity variables, the soft factors
$\bom{S}_a^{\pm,(0)} (\{\beta_m\},q) = \varepsilon_{\pm}\cdot \bom{J}^{a,(0)}(q)$
can be obtained by dotting polarization vectors of the soft gluon,
\begin{align}
\begin{split}
\varepsilon^{\nu}_+(q) 	
	& \mapsto \varepsilon^{\alpha \dot \alpha}_+(q)
	=  +\sqrt{2} \frac{\xi^\alpha \,  \lamt{q}^{\dot\alpha}}{\ab{q\xi}}\,,  \qquad
\varepsilon^{\nu}_-(q)   	
	 \mapsto \varepsilon^{\alpha \dot \alpha}_-(q)  
	= -\sqrt{2} \frac{\lam{q}^\alpha\, \widetilde{\xi}^{\dot\alpha}}{\sqb{q\xi}}\,,
\end{split}
\end{align}
into the soft current $\bom{J}^{a,(0)\,\mu}(q)$.
The arbitrary reference spinors $\xi^\alpha$ and $\widetilde{\xi}^{\dot\alpha}$
entering
\begin{align} \label{eq:treespinor}
  \frac{\varepsilon_+ \cdot p_i}{q\cdot p_i}
  - \frac{\varepsilon_+ \cdot p_j}{q\cdot p_j}
  = \frac{\sqrt{2}}{\ab{q\xi}}\left[\frac{\ab{\xi i}\ab{qj}
      - \ab{\xi j}\ab{qi}}{\ab{qi}\ab{qj}}\right] 
  = \sqrt{2} \frac{\ab{ij}}{\ab{iq}\ab{qj}}\,,
\end{align}
drop out in the second equality due to the Schouten identity.
The non-standard normalization of the color generators exactly 
cancels the $\sqrt{2}$ from the polarization vector, yielding
the tree-level soft factor in the helicity basis,
\begin{align} \label{eq:3}
\begin{split}
  \bom{S}_a^{{+},(0)}(\{\beta_m\}) &=
  + \frac{1}{2n} \sum_{i\neq j} (\bom{T}_i^a {-} \bom{T}_j^a)
  \frac{ \ab{ij}}{\ab{iq}\ab{qj}} \,, \\
  \bom{S}^{{-},(0)} (\{\beta_m\}) &=
  - \frac{1}{2n} \sum_{i \neq j} (\bom{T}_i^a {-} \bom{T}_j^a)
  \frac{ \sqb{ij}}{\sqb{iq}\sqb{qj} }\,.
\end{split}
\end{align}
At tree level, the soft factor is a sum over different gauge-invariant
dipole emissions. It is manifestly invariant under rescaling of individual
hard parton momenta, $p_m \to e^{t_m} p_m$, and therefore it only depends on
$\beta_m$, the angular direction of $p_m$.
Eq.~(\ref{eq:3}) is an important building block for
next-to-leading order QCD calculations, and for
leading and next-to-leading
logarithmic resummation.  Parity dictates that the emission
of a negative-helicity gluon can be obtained from the
positive-helicity case by the simple replacement $\ab{ab} \to  \sqb{ba}$.
Therefore, below we will present results only for the soft
factor associated with a positive-helicity gluon.
\begin{figure}[ht!]
    \centering
    \begin{subfigure}[c]{0.1\textwidth}
      \raisebox{0pt}{\includegraphics[scale=0.4]{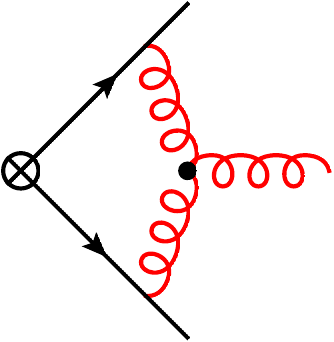}}
        \caption{}
        \label{fig:a}
    \end{subfigure}
    \qquad \qquad
    \begin{subfigure}[c]{0.1\textwidth}
      \includegraphics[scale=0.4]{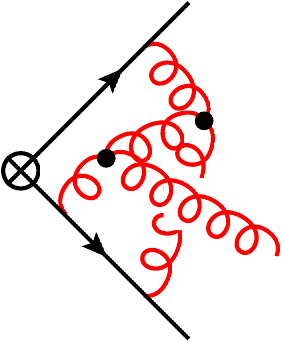}
        \caption{}
        \label{fig:b}
    \end{subfigure}
   \qquad
   \begin{subfigure}[c]{0.1\textwidth}
     \includegraphics[scale=0.4]{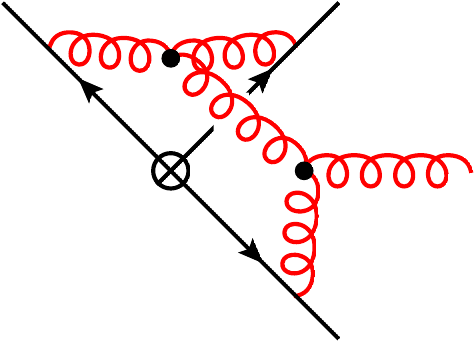}
        \caption{}
        \label{fig:c}
    \end{subfigure}
    \caption{Representative diagrams for the (a) one-loop, (b) two-loop dipole,
      and (c) two-loop tripole soft factors.}
\label{fig:1}
\end{figure}
%

%================================================
\vspace{-0pt}
\subsection{Soft factorization at one loop}
\label{subsec:soft_factorization_1L}\vspace{-4pt}
%================================================

At loop level, the soft factor receives corrections. The one-loop
soft factor was first obtained in ref.~\cite{Bern:1995ix}
(see also refs.~\cite{Bern:1998sc,Bern:1999ry})
for color-ordered amplitudes by taking the soft-gluon limit of the
one-loop splitting
amplitude~\cite{Bern:1994zx,Bern:1995ix,Bern:1998sc,Bern:1999ry}.
It was later re-derived for amplitudes in color space using
soft-gluon insertion techniques, \ie using the eikonal approximation for
both external and internal soft gluons~\cite{Catani:2000pi}. This computation
was performed in axial gauge. Since the soft factor is gauge invariant,
one can repeat the calculation in Feynman gauge, which is much easier
since there is only one non-vanishing diagram involved~\cite{Li:2013lsa},
shown in Fig.~\ref{fig:a}.
The result is again a sum over gauge-invariant dipole emissions
(see also ref.~\cite{Duhr:2013msa}):
\begin{align}
\label{eq:4}
\bom{S}^{+,(1)}_a
= & \, \frac{1}{2} C_1(\e) \sum_{i\neq j}  f_{aa_ia_j}  \bom{T}^{a_i}_i\, \bom{T}^{a_j}_j
\ V_{ij}^q\ \frac{\ab{ij}}{\ab{iq}\ab{qj}} \,,
\end{align}
where
\begin{align}
\label{eq:c1e}
C_1(\e) =     - \frac{e^{\e \gamma_E}}{\e^2}
			\frac{\Gamma^3(1-\e)\Gamma^2(1+\e)}{\Gamma(1-2\e)} 
	    = &  - \frac{1}{\e^2} - \frac{\zeta_2}{2} + \e \, \frac{7}{3} \zeta_3 + \e^2 \, \frac{39}{16} \zeta_4 + \mathcal{O}(\e^3), 
	    \\
\label{eq:16A}
V_{ij}^q = & \Vol{i}{j}^\e\,.    
\end{align}
We define the conventional Mandelstam variables
\begin{align}
  s_{ab} = \ab{ab}\sqb{ba} =
  - 2 \,\big| p_a \mcdot p_b \big| \exp(-\im \pi \lambda_{ab})  \,, 
\label{eq:16B}
\end{align}
with $\lambda_{ab}=1$ if $a$ and $b$ are both incoming or both outgoing,
and $0$ otherwise. Again, the one-loop soft factor is explicitly invariant
under rescaling of the hard parton momenta.
The analytic dependence on the parton momenta is fully captured by $V_{ij}^q$,
which is itself uniquely fixed by dimensional
analysis and rescaling symmetry, given that only two hard partons
can be involved in a non-factorizable one-loop diagram.
The analytic continuation of \eqn{eq:16A} to any kinematic region is
given by \eqn{eq:16B}, which is equivalent to letting
\vspace{-.3cm}
\begin{align}
  \log(-s_{ab}) \rightarrow \log|s_{ab}| - i\pi \Theta(s_{ab}),
\label{eq:lnsabcontinue}
\end{align}
where $\Theta(x)$ is the Heaviside theta function,
after expanding \eqn{eq:16A} in $\e$.

We emphasize that \eqn{eq:4} is the one-loop soft emission factor for
bare, unrenormalized amplitudes; one can also define an emission factor
for renormalized amplitudes but it will differ by a term proportional
to the tree emission factor multiplied by the beta function coefficient
$\beta_0$.  In the above unrenormalized case,
the one-loop soft factor is purely non-Abelian,
which is in agreement with the expectation \cite{Weinberg:1965nx} that
the soft photon limit in an Abelian gauge theory with no
massless charged fermions should be exact.  If there are massless
charged fermions, there is a correction proportional to $\beta_0$
to the soft emission factor for renormalized amplitudes, but
there is no correction at one loop to unrenormalized Abelian amplitudes.
\vspace{-.1cm}

%================================================
\vspace{-0pt}
\section{Soft factorization at two loops}
\label{sec:soft_factorization_2L}\vspace{-8pt}
%================================================

At two loops, the soft factor becomes substantially more involved. In the
large $N_c$ limit, also allowing for a large number of fermion flavors $N_f$,
the soft factor again factorizes into a sum of
gauge-invariant dipole emissions. The soft factor in this limit was first
obtained by taking the soft limit of the two-loop planar splitting
amplitude~\cite{Badger:2004uk}.  At higher orders in $\e$,
it has also been obtained by evaluating the soft limit of the integrals
entering the two-loop amplitude for $e^+e^- \to q\bar{q}g$~\cite{Duhr:2013msa},
and by direct calculation using eikonal techniques~\cite{Li:2013lsa}.

One of the main results of this work is the computation of the two-loop
non-planar contribution to the soft factor. 
In contrast to tree level and one loop, the soft factor contains
contributions from tripole emission as well as dipole emission:
\begin{align}
  \label{eq:6}
  \bom{S}^{+,(2)}_a =
  \frac{1}{2}\sum_{i\neq j} \bom{S}^{+,(2)}_{a, ij}
  - \frac{1}{4} \sum_{i\neq k \neq j}  \bom{S}^{+,(2)}_{a,  ikj}  \,.
\end{align}
Tripole emission is characterized by three separate hard legs, $i,j,k$.
Representative Feynman diagrams contributing to dipole and tripole emission
are depicted in fig.~\ref{fig:b} and \ref{fig:c}, respectively. The calculation is 
similar to the one in ref.~\cite{Li:2013lsa}. First, we generate all Feynman diagrams 
for the Wilson line matrix elements with a soft gluon using \texttt{QGRAF}~\cite{Nogueira:1991ex}. After performing color and 
Dirac algebra manipulations, the diagrams are passed to \texttt{LiteRed}~\cite{Lee:2012cn} for
Integration-By-Parts~\cite{Chetyrkin:1981qh} (IBP) reduction. Since the dipole integrals are already known~\cite{Li:2013lsa}, 
we focus on the tripole topology. Here, the IBP reduction leads to $8$ master integrals in total, which 
(owing to the nontrivial kinematic dependence) we solve using the method of differential equations~\cite{Bern:1993kr,Gehrmann:1999as,Henn:2013pwa}.

The result for the dipole emission is given by
\begin{align}
\label{eq:7}
\bom{S}^{+,(2)}_{a, ij} = &\
C_2(\e) \, f_{aa_i a_j} \bom{T}^{a_i}_i \, \bom{T}^{a_j}_j
\ \big(V_{ij}^q\big)^2\ \frac{\ab{ij}}{\ab{iq}\ab{qj}}\,,
\end{align}
where $C_2(\e)$ is a constant which can be expanded in $\e$:
\begin{align} \label{eq:c2e}
\begin{split}
C_2(\e) = &\ C_A \Bigg[
\frac{1}{2\e^4}{-}\frac{11}{12 \e^3} {+}
\frac{1}{\e^2}\left(\zeta_2 {-}\frac{16}{9}{-}\frac{\delta_R}{12} \right) {-}
\frac{1}{\e}\left(\frac{11\zeta_3}{6} {+}\frac{11\zeta_2 }{12}{+} \frac{181}{54}{+} \frac{2 \delta_R}{9}\right) \\
& \phantom{C_A \Bigg[} \hspace{1.55cm}
{+} \frac{7 \zeta_4}{8} {+} \frac{341 \zeta_3}{18} {-} \frac{16\zeta_2}{9} {-} 
    \frac{\delta_R \zeta_2}{12} {-} \frac{1037}{162} {-}\frac{35\delta_R}{54}
\Bigg]  \\
&
{+}\ T_R\, N_f \Bigg[
\frac{1}{3 \e^3} {+} \frac{5}{9 \e^2} {+} 
\frac{1}{\e}\left(\frac{\zeta_2}{3} +\frac{19}{27}\right)
- \frac{62 \zeta_3}{9} {+} \frac{5 \zeta_2}{9} + \frac{65}{81}
\Bigg] \\ 
&
{+}\ C_A\, N_s\Bigg[
\frac{1}{24\e^3} {+} \frac{1}{9 \e^2} {+} \frac{1}{\e}\left( \frac{\zeta_2}{24}{+} \frac{35}{108}\right)
- \frac{31 \zeta_3}{36} {+} \frac{\zeta_2}{9}{+} \frac{259}{324}
\Bigg] + \Ord(\e).
\end{split}
\end{align}
This result is presented for one adjoint vector, $N_f$ flavors of
fundamental fermions, and $N_s$ flavors of adjoint real scalars. The
parameter $\delta_R$ selects the regularization scheme, where $\delta_R=1$ 
corresponds to the 't Hooft-Veltman scheme~\cite{'tHooft:1972fi},
and $\delta_R=0$ corresponds to the
Four-Dimensional-Helicity scheme~\cite{Bern:1991aq,Bern:2002zk}
(which preserves supersymmetry here).
In SU$(N_c)$ gauge theory, $C_A = N_c$, and $T_R = 1/2$.
If we set $\delta_R=1$ and $N_s=0$, and take into account different
normalization conventions, then the result~(\ref{eq:c2e}) agrees
with ref.~\cite{Badger:2004uk}, and also with the terms through
$\Ord(\e^0)$ in refs.~\cite{Li:2013lsa,Duhr:2013msa}.  The latter
references give results valid to ${\cal O}(\e^2)$
and to all orders in $\e$, respectively, for the dipole terms in QCD.

The corresponding result in $\mathcal{N}=4$ SYM can be 
obtained by setting $T_R\to C_A/2$,
$N_f = 4$, and $N_s = 6$.
For $\delta_R=0$, the ${\cal N}=4$ SYM result has uniform transcendentality,
\begin{align}
C_2^{\mathcal{N}=4}(\e) = C_A \Bigg[ \frac{1}{2\e^4} + \frac{\zeta_2}{\e^2}
    - \frac{11\zeta_3}{6 \e} + \frac{7\zeta_4}{8} \Biggr] \,,
\end{align}
and it agrees with the soft ($z\to0$) limit
of the two-loop splitting amplitude in planar $\mathcal{N}=4$ SYM~\cite{ABDK}.

We note that the factor $C_2(\e)$ in \eqn{eq:c2e} contains no explicit
subleading-in-$N_c$ terms. However, there \emph{are} still non-planar
corrections to the dipole contribution.  In fact, both of the non-planar
diagrams in Figs.~\ref{fig:b} and \ref{fig:c} give rise to dipole
terms. In particular, the dipole contribution from fig.~\ref{fig:c} comes
from summing over color indices and applying the color Jacobi identity.
This statement is consistent with eq.~(\ref{eq:7}), in which the
color operator gives both leading- and subleading-color contributions when
acting on an amplitude with more than two colored hard partons.

%================================================
\vspace{-0pt}
\subsection{The tripole contribution}
\label{subsec:tripole}\vspace{-4pt}
%================================================

Before presenting the result for the tripole contribution
to the soft emission factor $\bom{S}^{+,(2)}_{a,  ikj}$,
it is useful to comment on the relevant kinematics.
For a given tripole of hard lines $(i,j,k)$, the soft factor can
only depend on $V_{ij}^q$ and rescaling-invariant cross ratios constructed
from $p_i,p_j,p_k$ and $q$. From the four momenta involved in the problem, 
one can form two independent cross ratios:
\begin{align}
  \label{eq:8}
  u^{ij}_k\ \equiv\ \frac{s_{ik} s_{jq}}{s_{ij} s_{kq}} \,,
  \qquad v^{ij}_k\ \equiv\ \frac{s_{jk} s_{iq}}{s_{ij} s_{kq}}  \,.
\end{align}
It turns out that the two-loop tripole emission contribution is a
complicated function of $u^{ij}_k$ and $v^{ij}_k$. In particular, 
it depends on the following K\"all\'en function:
\begin{align}
  \label{eq:9}
  \Delta^{ij}_k = \sqrt{ 1 - 2u^{ij}_k - 2 v^{ij}_k + \left(u^{ij}_k - v^{ij}_k\right)^2 } \,,
\end{align}
which gives rise to complications with the analytic continuation. To
rationalize the kinematics, we use the well-known parameterization
of the cross ratios:
\begin{align}
  \label{eq:10}
  u^{ij}_k\ =\ (1-z^{ij}_k) (1-\bar{z}^{ij}_k) \,, \qquad
  v^{ij}_k\ =\ z^{ij}_k \, \bar{z}^{ij}_k \,,
\end{align}
where $z^{ij}_k$ and  $\bar{z}^{ij}_k$ are cross ratios of spinor products,
\begin{align}
  \label{eq:11}
  z^{ij}_k \ =\ \frac{\ab{kj} \ab{iq}}{\ab{ij}\ab{kq}} \,, 
  \qquad
   \bar{z}^{ij}_k \ =\ \frac{\sqb{kj} \sqb{iq}}{\sqb{ij}\sqb{kq}} \,. 
\end{align}
Focusing on the holomorphic variables, it is useful to keep in mind that
$z^{ij}_k$ satisfies the Schouten identity,
\begin{align}
\label{eq:zjik}
  1- z^{ij}_k \ =\ 1 - \frac{\ab{kj} \ab{iq}}{\ab{ij}\ab{kq}}
  \ =\ \frac{\ab{ki} \ab{jq}}{\ab{ji}\ab{kq}}\ =\ z^{ji}_k\,.
\end{align}
The change of variables in eq.~\eqref{eq:10} sets 
\begin{align} \label{eq:imz}
\hspace{-.9cm}
\Delta^{ij}_k \, = \, z^{ij}_k {-}  \bar{z}^{ij}_k
\, = \, \frac{\sqb{ij}\ab{jk}\sqb{kq}\ab{qi}
    - \ab{ij}\sqb{jk}\ab{kq}\sqb{qi}}{\ab{ij}\sqb{ji} \, \ab{kq}\sqb{qk}}
\, = \, \frac{4 \im\, \varepsilon(p_i,p_j,p_k,q)}{\sab{ij} \sab{kq}}
\,,
\end{align}
where $\varepsilon(p_i,p_j,p_k,q)$ is the contracted Levi-Civita tensor.
Repeated use of the Schouten identity leads to four more relations,
\begin{align} \label{eq:zijk}
\hspace{-.9cm}
z^{kj}_i \, = \, \frac{1}{z^{ij}_k}\,, \qquad
z^{jk}_i \, = \, \frac{1 - z^{ij}_k}{-z^{ij}_k} \,, \qquad
z^{ik}_j \, = \, \frac{- z^{ij}_k}{1 - z^{ij}_k} \,, \qquad
z^{ki}_j \, = \, \frac{1}{1 - z^{ij}_k} 
\,,
\end{align}
which will be useful in summing the tripole soft factor.

The change of variables~(\ref{eq:10})
can be motivated by a stereographic projection, under which a
lightlike momentum is mapped to a point on the complex plane
(see \eg refs.~\cite{Lysov:2014csa,Almelid:2017qju}), and spinor products map
to differences of complex coordinates.
The SL$(2,\mathbb{C})$ invariance of the cross ratios
can then be used to map the points corresponding to $p_i,p_j,p_k$ to
$0,1,\infty$ on the complex plane, and the point corresponding to $q$ maps
to $z_k^{ij}$. In this chart, the limits $z^{ij}_k \to 0, 1, \infty$ correspond
to the collinear limits in which $q$ becomes parallel to
$p_i,p_j,p_k$, respectively.
On the real axis ($\mathrm{Im}(z^{ij}_k)=0$, \eqn{eq:imz})  
the volume element of $p_i,p_j,p_k$ and $q$ vanishes
($\varepsilon(p_i,p_j,p_k,q)=0$) so that the momenta lie in a
lower-dimensional subspace.  Parity exchanges
$\ab{ab} \leftrightarrow \sqb{ba}$, so it maps
$\Delta_{k}^{ij} \leftrightarrow - \Delta_{k}^{ij}$ and
is implemented as complex conjugation,
$z^{ij}_k \leftrightarrow \bar{z}^{ij}_k$.

We shall first present the result for the tripole emission soft factor in a
``Euclidean'' region\footnote{Depending on the momenta \emph{not} involved
  in the tripole emission, this region could still correspond to a
  physical scattering region.} 
in which all momenta participating in the tripole,
including the soft momentum, are either all incoming or all outgoing,
so that $u^{ij}_k>0$ and $v^{ij}_k>0$.
This region is denoted by $A_0$ in table~\ref{tab:1}.
Then we shall explain how to obtain the results in various physical regions
by analytic continuation, as summarized in table~\ref{tab:1}, where the phase
factors follow from eqs.~(\ref{eq:16B}) and (\ref{eq:8}).
In principle there are $2^4 = 16$ physical regions for each tripole,
since the three hard momenta $p_i,p_j,p_k$ and the soft momentum $q$
can be incoming or outgoing.  However, if \emph{all} incoming and outgoing momenta
are swapped with each other, the Mandelstam variables do not change.  Hence it
suffices to give the eight configurations with $q$ outgoing, as shown
in table~\ref{tab:1}.  Also, the cases $A_4$ to $A_7$, with
an odd number of incoming momenta and an odd number of outgoing ones,
always have one timelike and one spacelike invariant each
in the numerator and in the denominator of each cross ratio,
so the phases cancel and no continuation from $A_0$ is required.
Only cases $A_1$, $A_2$ and $A_3$ require a nontrivial continuation.
%
%%%%%%%% TABLE %%%%%%%%%
\begin{table}[h!]
  \centering
  \begin{tabular}{|c| l l | }
\hline
\hline
$A_0$: all outgoing  &     $u^{ij}_k = |u^{ij}_k|$  & $v^{ij}_k = |v^{ij}_k|$ 
\\
\hline
$A_1$: $j,k$ incoming, $i,q$ outgoing  &   $u^{ij}_k \to |u^{ij}_k|$ & $v^{ij}_k \to   |v^{ij}_k| e^{-2 \im \pi}$
\\
\hline
$A_2$: $i,k$ incoming, $j,q$ outgoing   &   $u^{ij}_k \to |u^{ij}_k| e^{-2 \im \pi}$ & $v^{ij}_k \to |v^{ij}_k|$
\\
\hline
$A_3$: $i,j$ incoming, $k,q$ outgoing   &   $u^{ij}_k \to |u^{ij}_k| e^{2 \im\pi}$ & $v^{ij}_k \to |v^{ij}_k| e^{2 \im\pi}$ 
\\
\hline
$A_4$: $i$ incoming, $j,k,q$ outgoing  &   $u^{ij}_k \to |u^{ij}_k|$ & $v^{ij}_k \to   |v^{ij}_k|$
\\
\hline
$A_5$: $j$ incoming, $i,k,q$ outgoing   &   $u^{ij}_k \to |u^{ij}_k|$ & $v^{ij}_k \to |v^{ij}_k|$
\\
\hline
$A_6$: $k$ incoming, $i,j,q$ outgoing   &   $u^{ij}_k \to |u^{ij}_k|$ & $v^{ij}_k \to |v^{ij}_k|$ 
\\
\hline
$A_7$: $i,j,k$ incoming, $q$ outgoing   &   $u^{ij}_k \to |u^{ij}_k|$ & $v^{ij}_k \to |v^{ij}_k|$ 
\\
\hline\hline
  \end{tabular}
  \caption{Rules for analytically continuing $u^{ij}_k$ and $v^{ij}_k$ from region $A_0$, also dubbed the ``Euclidean'' region, to other physical regions.  These rules can be obtained from eqs.~(\ref{eq:16B}) and (\ref{eq:8}). The rules are used below to continue the tripole emission terms to generic physical regions.}
  \label{tab:1}
\end{table}
%
%%%%%%%% END TABLE %%%%%%%%%
%

The tripole emission contribution can be decomposed into rational prefactors
multiplying transcendental functions.  The rational prefactors are
essentially the tree-level emission factors~(\ref{eq:treespinor}).
In principle there are three such factors, for the three pairs of legs
entering the tripole, $ij$, $jk$ and $ik$, but the Schouten identity
allows the $ij$ factor to be eliminated, leaving us with
\begin{align}
  \label{eq:12}
  \begin{split}
  \hskip -.1cm
    \bom{S}^{+,(2)}_{a, ikj} \, = & \, \xi_\e
	\big( V^q_{ij} \big)^2 
	f^{a a_k b} f^{b a_ia_j}  \bom{T}_i^{a_i} \bom{T}_j^{a_j} \bom{T}_k^{a_k} 
        \Bigg[ \frac{\ab{ik}}{\ab{iq}\ab{qk}} F(z^{ij}_{k},\e)
          \, - \, \frac{\ab{jk} }{\ab{jq}\ab{qk}} F(z^{ji}_{k},\e) \Bigg]\,
  \hskip -.4cm
 \end{split}
\end{align}
where the normalization factor $\xi_\e = 1 + \e^2 \zeta_2$ was
mistakenly omitted in an earlier version of this paper.
The transcendental function $F(z^{ij}_k,\e) \equiv F(z,\e)$ is
a function of $z$ and $\bar{z}$ on the complex plane.
Note that Bose symmetry under exchange of legs $i$ and $j$ is manifest
within a single tripole~(\ref{eq:12}), whereas symmetry under other exchanges,
such as $j$ and $k$, requires summing over different tripoles.
In the $A_0$ region, the soft factor cannot develop branch
cuts in the kinematic variables, unless one approaches the boundaries,
corresponding to Mandelstam variables vanishing, or equivalently, $u^{ij}_k$ and $v^{ij}_k$
going to $0$ or infinity.  These branch cuts are not in $z$, but
only in $|z|^2$ or $|1-z|^2$.   Such functions are called real analytic
functions, or single-valued functions of $z$ (with $\bar{z}=z^*$).
If they are polylogarithmic, then the first entries of their
symbol~\cite{Goncharov:2010jf}
can only be $z\bar{z}$ or $(1-z)(1-\bar{z})$~\cite{Gaiotto:2011dt}.

It turns out that through $\Ord(\e^0)$, the tripole emission contribution
can be fully described by a special class of such real analytic functions,
called single-valued harmonic
polylogarithms~(SVHPLs)~\cite{Brown:SVHPLs,Dixon:2012yy}. 
Interestingly, such functions also appear in the context of the soft-gluon
anomalous dimension at three loops~\cite{Almelid:2015jia,Almelid:2017qju}.
In that case, SVHPLs with transcendental weight up to 5 can appear,
while here we encounter at most weight 4.
The SVHPLs related to our problem are denoted by $\cL_{\vec{w}}(z,\bar{z})$,
where $\vec{w}$ is a binary string of 0's and 1's of length $|w|$ equal to the
\emph{weight}.  The $\bar{z}$ dependence will often be implicit below.
They are defined by the differential equations
\begin{align}
  \label{eq:13}
  \frac{\partial}{\partial z} \cL_{0,\vec{w}}(z,\bar{z})
  = \frac{\cL_{\vec{w}}(z,\bar{z})}{z}\,,
\qquad
\frac{\partial}{\partial z} \cL_{1,\vec{w}}(z,\bar{z})
= \frac{\cL_{\vec{w}}(z,\zb)}{1-z}\,,
\end{align}
subject to the constraints of single-valuedness and
\begin{align*}
  \cL_{0^n}(z,\bar{z}) =  \frac{1}{n!} \log^n|z|^2 \,, \ \ n\geq0, \quad\ 
  \cL_1(z,\bar{z}) = - \log|1-z|^2 \,, \quad\  
  \lim_{z\to 0} \cL_{\vec{w} \neq 0^n}(z,\bar{z}) = 0 \,.
\end{align*}
An explicit construction of SVHPLs in terms of the more familiar
harmonic polylogarithms~(HPLs)~\cite{Remiddi:1999ew} is given in
ref.~\cite{Dixon:2012yy}. SVHPLs have also been implemented in 
the \texttt{Mathematica} package \texttt{PolyLogTools} \cite{Duhr:2019tlz} 
which allows easy manipulation of expressions containing SVHPLs.

As mentioned in the introduction, the tripole contribution to the two-loop 
soft factor contains only soft gluons and is therefore the same in any gauge theory, including $\N=4$ SYM. Thanks to the uniform transcendentality property of $\mathcal{N}=4$ SYM,
we expect $F(z,\e)$ to be a function of uniform weight
$4$~(counting $\e$ with transcendental weight $-1$), and indeed it is:\footnote{Occasionally, we suppress the explicit functional dependence on $\zb$. When the explicit dependence is required, {\it e.g.}~in order to specify a particular analytic continuation where $\zb \neq z^*$, we write both arguments. We hope our shortened notation does not cause any confusion.}
\begin{align}
\label{eq:soft_tripole_func}
\begin{split}
  F(z,\e) &=  \frac{1}{\e^2} \cL_0 \cL_1
  		+\frac{1}{3 \e}\Big( \cL_0^2 \cL_1-2 \cL_0 \cL_1^2 \Big) \\
 &\hskip0.2cm
- \cL_1 \left( \frac{2}{9} \cL_0^3 + \frac{1}{3} \cL_0^2 \cL_1
+ \frac{13}{18} \cL_0 \cL_1^2 + \frac{7}{12} \cL_1^3 \right)
+ 2 \cL_{1,0,1,0} 
\\
&\hskip0.2cm + \frac{4}{3} \bigg[  2 \Big( \cL_{0,0,0,1} + \cL_{0,0,1,0}\Big)
    + \cL_{0,0,1,1} - \cL_{0,1,1,1}  - \cL_{1,0,1,1} - \cL_{1,1,0,0} \bigg]
\\
&\hskip0.2cm
+ 2 \zeta_2 \Bigl( 2 {\cal L}_{0,1} - {\cal L}_0 {\cal L}_1 \Bigr)
+ \frac{40}{3} \zeta_3 \, \cL_1  \ +\ \Ord(\e) \,,
\end{split}
\end{align}
where $\cL_{\vec{w}} = \cL_{\vec{w}}(z,\bar{z})$.
Explicit expressions for $F(z,\e)$ in terms of
harmonic polylogarithms in $z$ and $\bar{z}$, and
in terms of $G$ functions, as well as its symbol,
are provided in an ancillary file.

\vfill\eject

%================================================
\vspace{-0pt}
\subsection{Alternate representation of tripole contribution}
\label{subsec:tripolealt}\vspace{-4pt}
%================================================

In this subsection we provide another representation of
the contributions of a single tripole $\{i,j,k\}$, which is
more convenient for many purposes, including
describing the analytic continuation into regions $A_1$, $A_2$, and $A_3$.
It turns out that the single term $\bom{S}^{+,(2)}_{a, ikj}$ given
in \eqn{eq:12} becomes {\it ambiguous} in the kinematic regions
$A_1$, $A_2$, and $A_3$.  The ambiguities cancel, as they must,
after summing over all six permutations of $i,j,k$ corresponding
to a given tripole.

Thus we wish to rewrite the second term of \eqn{eq:6} as
a sum over distinct tripoles, labelled by the unordered set $\{i,j,k\}$:
\be
- \frac{1}{4} \sum_{i\neq k \neq j}  \bom{S}^{+,(2)}_{a,  ikj}
\ =\ - \frac{1}{4} \sum_{\substack{ \text{tripoles}\\ \{i,j,k\}}} \bom{S}^{+,(2)}_{a, \{i,j,k\}} \,,
\label{tripolesumALT}
\vspace{-.5cm}
\ee
where
\begin{align}
\label{eq:tripolesum}
\hspace{-1.1cm}
\bom{S}^{+,(2)}_{a, \{i,j,k\}} 
= &\  2 \Bigl( \bom{S}^{+,(2)}_{a,  ikj} + \bom{S}^{+,(2)}_{a,  kji}
         + \bom{S}^{+,(2)}_{a,  jik} \Bigr)
\nn\\
 = &\ 
  2 \, \bom{T}^{a_i}_i \bom{T}^{a_j}_j \bom{T}^{a_k}_k \, \bigg\{ 
 \frac{\ab{ik}}{\ab{iq}\ab{qk}} \, \xi_\e \, (V_{ik}^q)^2 
 \Bigl[ f^{a a_j b} f^{b a_ia_k}  D_1(z,\bar{z})	
       + f^{a a_i b} f^{b a_ka_j}  D_2(z,\bar{z}) 
 \Bigr]
 \nn \\
& \hskip4.9cm +\ \{i \leftrightarrow j\} \qquad\qquad  \bigg\} \,,
\hspace{-2.5cm}
\end{align}
with
\bea
D_1(z,\bar{z}) &=& u^{-2 \e}\, F(z,\bar{z})
+  F\left(\frac{{-}z}{1{-}z},\frac{{-}\bar{z}}{1{-}\bar{z}}\right)\,,
\label{D1def}\\
D_2(z,\bar{z}) &=& u^{-2 \e}\, F(z,\bar{z})
- \left(\frac{u}{v}\right)^{-2 \e} \biggl[
    F\Bigl(\frac{1}{z},\frac{1}{\bar{z}}\Bigr)
  - F\Bigl(\frac{1{-}z}{{-}z},\frac{1{-}\bar{z}}{{-}\bar{z}}\Bigr) \biggr]\,,
\label{D2def}
\eea
and $z = z_k^{ij},\ 1{-}z = z_k^{ji}$ via the definitions in eqs.~(\ref{eq:11}) and~(\ref{eq:zjik}).

In order to obtain \eqn{eq:tripolesum} we used the manifest symmetry of
$\bom{S}^{+,(2)}_{a, ikj}$ under exchange of legs $i$ and $j$,
the Schouten relation between tree-level eikonal factors,
and the Jacobi relation for the gauge-theory structure constants,
thereby rearranging the sum into a minimal number of transcendental functions,
$D_1$ and $D_2$.  This rearrangement is important because $D_1$ and $D_2$
should have unambiguous analytic continuations into regions $A_1$, $A_2$ and
$A_3$. (As we will explain shortly, using the symmetry of the tripole
formula~(\ref{eq:tripolesum}), it is sufficient to discuss only the
$A_1$ discontinuity.) In contrast, the analytic continuation of $F$
has ambiguities that only cancel due to identities among its
rational prefactors. 

The functions $D_1$ and $D_2$ can be computed from $F$ in
\eqn{eq:soft_tripole_func}, after transforming the 
arguments\footnote{In \texttt{PolyLogTools} \cite{Duhr:2019tlz} this can 
be achieved with the ``ToFibrationBasis'' command. } of
the SVHPLs back to a uniform argument $(z,\bar{z})$.  The results are:
\bea 
D_1(z) &=& - \frac{1}{\e^2} (\cL_{1})^2 - \frac{1}{\e} (\cL_{1})^3
           - \frac{7}{12} (\cL_{1})^4
           + 4 \cL_{1,0,1,0} + 2 \cL_{1,0,1,1} + 2 \cL_{1,1,1,0} \,,
 \label{D1value}\\ 
D_2(z) &=& \frac{1}{\e^2} \cL_{0} \cL_{1} + \frac{1}{\e} \cL_{0} (\cL_{1})^2
          + \frac{2}{3} \cL_{0} (\cL_{1})^3
          + 6 \, \zeta_2 \, \bigl( \cL_{0,1} - \cL_{1,0} \bigr) 
\nn\\
&& \ +\ 2 \bigl( \cL_{0,0,0,1} - \cL_{0,0,1,0}
              + \cL_{0,1,0,0} + \cL_{0,1,0,1} - \cL_{1,0,0,0} \bigr)
\,. \label{D2value}
\eea
For the $i \lr j$ term in \eqn{eq:tripolesum}, it is convenient
to have the same functions with argument $1-z$, expressed in terms
of SVHPLs with argument $z$:
\bea 
D_1(1-z) &=& - \frac{1}{\e^2} (\cL_{0})^2 + \frac{1}{\e} (\cL_{0})^3
       - \frac{7}{12} (\cL_{0})^4
       + 4 \cL_{0,1,0,1} + 2 \cL_{0,1,0,0} + 2 \cL_{0,0,0,1}
\nn\\&&\hskip0cm\null
       + 8 \zeta_3 \, \cL_{0} \,,
 \label{D1omzvalue}\\ 
D_2(1-z) &=& \frac{1}{\e^2} \cL_{0} \cL_{1} - \frac{1}{\e} (\cL_{0})^2 \cL_{1}
          + \frac{2}{3} (\cL_{0})^3 \cL_{1}
          + 6 \, \zeta_2 \, \bigl( \cL_{1,0} - \cL_{0,1} \bigr)
          - 8 \zeta_3 \, \cL_{1} 
\nn\\
&& \ +\ 2 \bigl( \cL_{1,1,1,0} - \cL_{1,1,0,1}
              + \cL_{1,0,1,1} + \cL_{1,0,1,0} - \cL_{0,1,1,1} \bigr)
\,. \label{D2omzvalue}
\eea

It is possible to write $D_1$ and $D_2$ explicitly in terms of classical
polylogarithms. 
However, the results are not particularly compact, and
the single-valuedness is not particularly manifest:

\bea
D_1(z) &=&  - \, \frac{\log^2|1-z|^2}{\e^2} + \frac{\log^3|1-z|^2}{\e}
\nn\\&&\hskip0cm\null
+ 8 \, \biggl[ {\rm Li}_4(z) - {\rm Li}_4(\bar{z})
  + {\rm Li}_4\Bigl(\frac{-z}{1-z}\Bigr)
  - {\rm Li}_4\Bigl(\frac{-\bar{z}}{1-\bar{z}}\Bigl) \biggr]
\nn\\&&\hskip0cm\null
+ 4 \, \Bigl( \log|1-z|^2 - 2 \log|z|^2 \Bigr)
    \biggl[ {\rm Li}_3(z) + {\rm Li}_3\Bigl(\frac{-z}{1-z}\Bigl) \biggr]
\nn\\&&\hskip0cm\null
+ 4 \,\log|1-z|^2 \, \biggl[ {\rm Li}_3(\bar{z})
      - {\rm Li}_3\Bigl(\frac{-\bar{z}}{1-\bar{z}}\Bigl) \biggr]
+ 2 \, \Bigl( {\rm Li}_2(z) - {\rm Li}_2(\bar{z}) \Bigr)^2
\nn\\&&\hskip0cm\null
+ \Bigl( 4 \log(1-z) \log|z|^2 - \log^2|1-z|^2 \Bigr)
  \Bigl( {\rm Li}_2(z) - {\rm Li}_2(\bar{z}) \Bigr)
\nn\\&&\hskip0cm\null
  + \log(1-z) \log\Bigl(\frac{1-z}{1-\bar{z}}\Bigr) \, \log|z|^2 \, \log|1-z|^2
\nn\\&&\hskip0cm\null
  - \frac{1}{12} \Bigl( 11 \, \log(1-z) + 3 \, \log(1-\bar{z}) \Bigr)
     \log^3|1-z|^2 \,,
\label{D1polylog}
\eea
\vskip -.5cm
\bea
D_2(z) &=& - \, \frac{\log|z|^2 \, \log|1-z|^2}{\e^2}
+ \frac{\log|z|^2 \, \log^2|1-z|^2}{\e}
\nn\\&&\hskip0cm\null
+ 4 \Biggl\{ 3 \, \Bigl( {\rm Li}_4(z) - {\rm Li}_4(\bar{z}) \Bigr)
      - {\rm Li}_4\Bigl(\frac{-z}{1-z}\Bigr)
      + {\rm Li}_4\Bigl(\frac{-\bar{z}}{1-\bar{z}}\Bigl)
\nn\\&&\hskip0.8cm\null
      + {\rm Li}_4(1-z) - {\rm Li}_4(1-\bar{z})
      + \Bigl( \log|1-z|^2 - 2 \, \log|z|^2 \Bigr)  {\rm Li}_3(z)
\nn\\&&\hskip0.8cm\null
      + \log|z|^2 \, \biggl[ {\rm Li}_3(\bar{z})
           - {\rm Li}_3\Bigl(\frac{-\bar{z}}{1-\bar{z}}\Bigl) \biggl]
        \Biggr\}
\nn\\&&\hskip0cm\null
+ \Bigl( {\rm Li}_2(z) - {\rm Li}_2(\bar{z}) \Bigr)^2
+ 2 \, \log|z|^2 \,  \Bigl( \log|z|^2 - \log(1-\bar{z}) \Bigr)
    \, \Bigl( {\rm Li}_2(z) - {\rm Li}_2(\bar{z}) \Bigr)
\nn\\&&\hskip0cm\null
- \frac{1}{12} \, \log|1-z|^2 \, \log^3 \Bigl(\frac{1-z}{1-\bar{z}}\Bigr)
+ \frac{1}{2} \, \log z \, \log|1-z|^2
              \, \log^2 \Bigl(\frac{1-z}{1-\bar{z}}\Bigr)
\nn\\&&\hskip0cm\null
+ \frac{1}{3} \, \log^3 |z|^2  \, \log\Bigl(\frac{1-z}{1-\bar{z}}\Bigr)
+ \frac{1}{12} \, \Bigl[ \log\Bigl(\frac{z}{\bar{z}}\Bigr)
                       - \log\Bigl(\frac{1-z}{1-\bar{z}}\Bigr) \Bigr]
                   \, \log^3 |1-z|^2
\nn\\&&\hskip0cm\null
- \frac{7}{12} \, \log|z|^2 \, \log^3 |1-z|^2
\nn\\&&\hskip0cm\null
+ \zeta_2 \, \Biggl\{ 6 \, \biggl[
                2 \, \Bigl( {\rm Li}_2(z) - {\rm Li}_2(\bar{z}) \Bigr)
              + \log|z|^2 \log\Bigl(\frac{1-z}{1-\bar{z}}\Bigr) \biggr]
\nn\\&&\hskip0.8cm\null
           - 2 \Bigl( \log^2(1-z) - \log^2(1-\bar{z}) \Bigr) \Biggr\}
\nn\\&&\hskip0cm\null
- 4 \, \zeta_3 \, \log|1-z|^2 \,.
\label{D2polylog}
\eea
\vskip -.1cm \noindent
These versions are valid for $0 < {\rm Re} \, z < 1$,
although they can also be extended outside this range.
Note that in order to evaluate $\log\left(\frac{1-z}{1-\zb}\right)$ numerically, 
depending on the computer algebra system, one might have to write this 
as $\log(1-z) - \log(1-\zb) $. For example, \texttt{Mathematica} always assigns 
a phase between $-\pi$ and $+\pi$ to $(1-z)/(1-\zb)$, which can be different from 
the actual phase.

For complex $z$, the $D_i$ functions are generically complex.
On the Euclidean sheet, they obey a reality condition:
When $z$ is complex conjugated, the functions get complex conjugated,
\be
D_i(\bar{z},z) = \overline{D_i(z,\bar{z})} \,.
\label{cxconjrel}
\ee
Interestingly, the imaginary part of $D_1(z,\bar{z})$
vanishes identically on the circle of radius 1 centered at $z=1$,
which has $|z|^2 = 2 \, {\rm Re}\,z$, or $\bar{z} = -z/(1-z)$.
One can see manifestly that the ${\rm Li}_4$ terms in
\eqn{D1polylog} vanish on this circle, and using $\log|1-z|^2=0$
it is easy to see that the rest of the terms are real.
The same is not true for $D_2$.

In the case that $z$ is real
(when the volume element of $p_i,p_j,p_k$ and $q$ vanishes),
$D_1$ and $D_2$ simplify considerably, to 
\bea
D_1(\bar{z}=z) &=&
- \frac{4}{\e^2} \, \log^2|1-z| + \frac{8}{\e} \, \log^3|1-z|
- \frac{28}{3} \, \log^4|1-z|
\nn\\ &&\hskip0.0cm\null
- 16 \, \log\Bigl|\frac{z}{1-z}\Bigr| \, {\rm Li}_3(z)
- 16 \, \log|z| \, {\rm Li}_3\Bigl(\frac{-z}{1-z}\Bigl) \,,
\label{eq:D1realz}\\
D_2(\bar{z}=z) &=&
- \frac{4}{\e^2} \, \log|z| \, \log|1-z|
+ \frac{8}{\e} \, \log|z| \, \log^2|1-z|
- \frac{28}{3}  \, \log|z| \, \log^3|1-z|
\nn\\ &&\hskip0.0cm\null
- 8 \, \log\Bigl|\frac{z}{1-z}\Bigr| \, {\rm Li}_3(z)
- 8 \, \log|z| \, {\rm Li}_3\Bigl(\frac{-z}{1-z}\Bigl)
- \, 8 \, \zeta_3 \, \log|1-z| \,,
\label{eq:D2realz}
\eea
where for $z > 1$ one should replace
\bea
{\rm Li}_3(z) &\to& {\rm Li}_3\Bigl(\frac{1}{z}\Bigr)
                 - \frac{1}{6} \log^3 z + 2 \zeta_2 \log z \,,
\label{Li3zreplace}\\
{\rm Li}_3\Bigl(\frac{-z}{1-z}\Bigl) &\to& 
- \, {\rm Li}_3(1-z) - {\rm Li}_3\Bigl(\frac{1}{z}\Bigr)
+ \frac{1}{6} \, \log^3 z + \frac{1}{6} \, \log^3(z-1)
\nn\\ &&\hskip0.0cm\null
- \frac{1}{2} \, \log z \, \log^2(z-1)
+ \zeta_2 \log\Bigl(\frac{z-1}{z^2}\Bigr) + \zeta_3 \,,
\label{Li3zomzreplace}
\eea
so that the expressions remain manifestly real.
While \eqns{eq:D1realz}{eq:D2realz}
contain only trilogarithms (${\rm Li}_3$)
and lower-weight functions,
it is clear from \eqns{D1polylog}{D2polylog} that the parts of $D_1$
and $D_2$ that are odd under $z \leftrightarrow \bar{z}$
contain ${\rm Li}_4$ as well,
but those parts vanish on the real line $z=\bar{z}$.
It is also interesting that the polylogarithmic parts of $D_1$ and
$2\,D_2$ are identical on the real axis.

%================================================
\vspace{-0pt}
\subsection{Analytic continuation}
\label{subsec:analytic_cont}\vspace{-4pt}
%================================================

So far we have restricted our discussion to the
$A_0$ region. To derive phenomenologically relevant results, it is
necessary to perform an analytic continuation of the soft factor. The
analytic continuation of the dipole terms is trivial and is
completely specified by eqs.~\eqref{eq:16A} and \eqref{eq:16B}.
However, the analytic continuation for the tripole contribution
is more involved. If zero, two, or three hard legs are outgoing 
(in addition to the outgoing soft gluon $q$), table~\ref{tab:1} shows
that no analytic continuation from $A_0$ is required for either $F$ or $D_i$. 
(Note that the $V_{ij}^q$ prefactors can acquire phases from \eqn{eq:16B}.)
In the $A_0$ region, the functions $F$ and $D_i$ are real on the real axis
and they complex conjugate when $z$ does; see \eqn{cxconjrel}.
As mentioned before, all cases where $q$ is incoming can be obtained
from the ones in table~\ref{tab:1} by exchanging all incoming and all outgoing
momenta. 

For the three Minkowski regions $A_1$, $A_2$ and $A_3$ described 
in table~\ref{tab:1}, the analytic continuation of the tripole 
emission term is nontrivial. 
However, using the symmetric tripole formula $\bom{S}^{+,(2)}_{a,\{i,j,k\}}$,~\eqn{eq:tripolesum}, it suffices
to give the $A_1$ discontinuity. In order to see this, note that the tripole $\{i,j,k\}$
is unordered. Therefore, if the soft momentum $q$ is outgoing
and exactly one of the three hard legs in the tripole is outgoing, we label
that leg by $i$. According to table~\ref{tab:1}, such a tripole
should be evaluated in region $A_1$, which we direct our attention to in the following. 

The difficulty in the analytic continuation of the tripole terms comes from the fact that $z^{ij}_k$ and $\bar{z}^{ij}_k$
contain square roots of $u^{ij}_k$ and $v^{ij}_k$, see the quadratic relations in eq.~\eqref{eq:10}.
To determine the analytic continuation of the SVHPLs entering the $D_i$ functions,
we use a bottom-up approach~(in the sense of transcendental weight).
We focus on the tripole labelled by $\{i,j,k\}$
and drop the particle indices $i,j,k$ for now.
At weight $1$ there are only two SVHPLs, due to the first-entry condition
mentioned above:
\begin{align}
  \label{eq:17}
  \cL_0(z) = \log (z\bar{z}) = \log v \,, \qquad
  \cL_1(z) = - \log[(1-z)(1-\bar{z})] = - \log u \,,  
\end{align}
whose analytic continuation properties are specified by eq.~\eqref{eq:16B}
and are summarized in table~\ref{tab:1}.
Starting from weight $1$, we can build the analytic continuation for
weight $2$ SVHPLs by requiring consistency with the differential equations.
As an example, consider $\cL_{0,1}(z)$, which in terms of
logarithms and polylogarithms is given by
\begin{align}
  \label{eq:18}
  \cL_{0,1}(z) = - \log( 1 - \bar{z} )  \log (z \bar{z})
  + \Li_2 (z) - \Li_2(\bar{z}) \,.
\end{align}
While it is not so difficult to compute the discontinuity of
$\cL_{0,1}(z)$ directly, it is even easier to compute the discontinuity
of its \emph{derivative}:
\begin{align}
  \label{eq:19}
  \partial_z \cL_{0,1}(z) = \frac{\cL_1(z)}{z} \,, \qquad
  \partial_{\bar z} \cL_{0,1}(z) = \frac{\cL_0(z)}{ 1 - \bar{z} }  \,.
\end{align}
For a given region, we first tabulate the discontinuities of
$\cL_0$ and $\cL_1$.  We then use the fact that the operations
of taking the discontinuity and taking the derivative commute.
We take the discontinuity of eq.~\eqref{eq:19} and then integrate up
the right-hand-side, to get the discontinuity of $\cL_{0,1}(z)$,
up to an additive constant. To determine the constant, we
can work near the point where the analytic continuation is being
performed. There, the function can only involve logarithms
plus irrelevant power corrections and it is easy to analytically continue.

For example, suppose we want to continue all the functions into the $A_1$
region.  We can deform $z \to z\, e^{-2\pi \im}$ around the origin,
keeping $\bar{z}$ constant.\footnote{Since $\cL_{0,1}(z)$ is a
  real-analytic function, we can freely split the analytic continuation
  in terms of $z$ and $\bar{z}$. Equivalently, we could have chosen the
  symmetric deformation $z \to z\, e^{-\im \pi}, \bar{z} \to\bar{z}\, e^{-\im \pi}$.}
The discontinuities in $\cL_0$ and $\cL_1$ are
\begin{align}
  \label{eq:20A}
\begin{split}
  \underset{z\to z\, e^{-2\pi \im }}{\mathrm{disc}}
  \,\big[ \cL_{0}(z,\bar{z}) \big]
  \ \equiv\ &  \big[ \cL_{0}(z e^{-2\pi \im},\bar{z}) -  \cL_{0}(z,\bar{z})\big]
 \ =\  - 2 \im \pi \,, \\
  \underset{z\to z\, e^{-2\pi \im }}{\mathrm{disc}}
  \,\big[ \cL_{1}(z,\bar{z}) \big]
  \ \equiv& \ \big[ \cL_{1}(z e^{-2\pi \im},\bar{z}) -  \cL_{1}(z,\bar{z})\big]
  \ =\  0 \,,
\end{split}
\end{align}
where we have explicitly written two arguments in the (poly)logarithms
to indicate that we do not enforce $\bar{z} = z^*$ in the analytic
continuation. Now we plug \eqn{eq:20A} into the discontinuity of
\eqn{eq:19} and integrate up, 
\begin{align}
  \label{eq:20}
  \underset{z\to z\, e^{-2\pi \im }}{\mathrm{disc}}
  \,\big[ \cL_{0,1}(z,\bar{z}) \big]\ =\ 2 \im \pi \log (1 - \bar{z}) \,.
\end{align}
The constant can be fixed at $z=\bar{z}=0$, where
$\cL_{0,1} \to 0 \times \log(z\bar{z})$, and so the discontinuity must
vanish at that point. Note that the discontinuity is no longer a
single-valued function.  Using this bottom-up approach we can determine
the discontinuity (and therefore the analytic continuation) of not just
$\cL_{0,1}$ but of all the SVHPLs, iteratively to higher weight.

We shall provide the discontinuities $\mathrm{disc}_{A_1} D_i$ of the
functions $D_i$ to go from region $A_0$ to regions $A_1$.
The function itself is given by
\be
D_i(z,\bar{z})|_{A_1}
\ =\ D_i(z,\bar{z})|_{A_0} + \mathrm{disc}_{A_1} D_i(z,\bar{z}),
\label{fnvsdisc}
\ee
where
\be
\mathrm{disc}_{A_1} D_i(z,\bar{z})\ =\
  \underset{z\to z\, e^{-2\pi \im }}{\mathrm{disc}}
  \left[ D_i(z,\bar{z}) \right]\,.
\label{discA1}
\ee
The result for the discontinuity needed to move $D_1(z)$ and $D_2(z)$ from
region $A_0$ to region $A_1$ can be expressed in terms of
classical polylogarithms:
\bea
\mathrm{disc}_{A_1} D_1(z)
&=&  2 \im \pi \biggl\{
8 \Bigl[ \text{Li}_3(z) + \text{Li}_3\Big(\frac{-z}{1-z}\Big) \Bigr]
\nn\\&&\hskip0.8cm\null
- \log(1-z) \Bigl[ 4 \Bigl( {\rm Li}_2(z) - {\rm Li}_2(\bar{z}) \Bigr)
   + \log^2(1-z) - \log^2(1-\bar{z}) \Bigr]  
\biggr\} \,. 
\nn\\&&\hskip0.8cm\null~~~
\label{discA1D1}
\eea
In this form, $\mathrm{disc}_{A_1} D_1(z)$ manifestly has no branch cuts for
${\rm Re} \, z < 1$.  For ${\rm Re} \, z > 1$ it is not manifest,
but one can check that as $z$ approaches the real axis, the imaginary
part in $\mathrm{disc}_{A_1} \big[D_1(z)/(2 \im \pi)\big]$ cancels,
and the result is {\it unambiguous}, and equal to:
\bea
\mathrm{disc}_{A_1} D_1(z)\bigl|_{z>1,\ {\rm real}}
&=&  2 \im \pi \times 8 \Bigl\{
  - \text{Li}_3(1-z) - \frac{1}{2} \log z \Bigl( \log^2(z-1) + \pi^2 \Bigr) 
\nn\\&&\hskip1.6cm\null
  + \frac{1}{6} \log^3(z-1) + \zeta_2 \log(z-1) + \zeta_3 \Bigr\} \,.
\label{discA1D1zgt1}
\eea
Similarly, the $A_1$ discontinuity of the $D_2(z)$ function is
\bea
\mathrm{disc}_{A_1} D_2 (z) &=& 
2 \im \pi  \Bigg\{ 
\frac{ \log|1-z|^2}{\e^2}-\frac{ \log^2 |1-z|^2}{\e}
+ 8 \text{Li}_3(z) - 4 \text{Li}_3(\bar{z})
+ 4 \text{Li}_3\Big( \frac{-\zb}{1-\zb} \Big)   
\nn\\&&\hskip0.8cm\null
- 2 \Big[\text{Li}_2(z)-\text{Li}_2(\bar{z} ) \Big]
   \Big[2 \log |z|^2 - \log (1-\zb) \Big]
+ 2 \zeta_2 \log \left(\frac{1-z}{1-\zb} \right)
\nn\\&&\hskip0.8cm\null
- \log\left( \frac{1-z}{1-\zb} \right) \, \log^2 |z|^2
+ 2 \log (1-z) \log (1-\zb) \, \log |1-z|^2
\nn\\&&\hskip0.8cm\null
+ \frac{2}{3} \log^3(1-z)  
\Bigg\}
\nn\\&&\hskip0.0cm\null
-4 \pi ^2  \left\{
2 \Big[ \text{Li}_2(z) - \text{Li}_2(\bar{z} ) \Big]
+ \log \left( \frac{1-z}{1-\bar z} \right) \, \log |z|^2 \right\} \,.
\label{discA1D2}
\eea
This result also has a well-defined limit in the region where
$z=\zb \in (1, \infty)$,
\bea
\mathrm{disc}_{A_1} D_2(z)\bigl|_{z>1,\ {\rm real}}
&=&  4 \im \pi \biggl\{
     \frac{\log(z-1)}{\e^2} - \frac{2 \log^2(z-1)}{ \e} 
     - 2 \text{Li}_3 (1-z) + \frac{8}{3} \log^3(z-1)
\nn\\&&\hskip0.8cm\null
     - \log z \left(\log^2(z-1) + \pi^2 \right)
     + 2 \zeta_2 \log(z-1) + 2 \zeta_3 \biggr\} \,.
\label{discA1D2zgt1}
\eea
We also need the $A_1$ discontinuity of $D_1(1-z)$,
\bea
\mathrm{disc}_{A_1} D_1(1-z)
&=&  2 \im \pi \biggl\{
\frac{2 \log|z|^2}{\e^2}
- \frac{3}{\e} \Bigl( \log^2|z|^2 - 8 \zeta_2 \Bigr)
+ 4 {\rm Li}_3(z) + 4 {\rm Li}_3(\bar{z})
\nn\\&&\hskip0.8cm\null
+ 8 {\rm Li}_3\Big( \frac{-\zb}{1-\zb} \Big)
- 2 \Big[\text{Li}_2(z)-\text{Li}_2(\bar{z} ) \Big]
 \Big[ \log|z|^2 - 2 \log(1-\zb) \Big]
\nn\\&&\hskip0.8cm\null
- \frac{4}{3} \log^3(1-\zb) + \log(1-\zb) \, \log^2|z|^2
+ \frac{7}{3} \log^3|z|^2
\nn\\&&\hskip0.8cm\null
- 8 \zeta_2 \Big[ 7 \log|z|^2 + \log(1-\zb) \Big]
- 8 \zeta_3 \biggl\}
\nn\\&&\hskip0.0cm\null
- 4 \pi^2 \biggl\{ - \frac{1}{\e^2} + \frac{3}{\e} \log|z|^2
- \frac{7}{2} \log^2|z|^2 + {\rm Li}_2(z) - {\rm Li}_2(\zb)
\nn\\&&\hskip1.4cm\null
- \log(1-\zb) \, \log|z|^2 + 14 \zeta_2 \biggr\} \,.
\label{discA1D1omz}
\eea
For $z=\zb \in (1, \infty)$ it becomes
\bea
\mathrm{disc}_{A_1} D_1(1-z)\bigl|_{z>1,\ {\rm real}}
&=& 2 \im \pi \biggl\{
\frac{4 \log z}{\e^2} - \frac{12}{\e} \Bigl[ \log^2 z - 2 \zeta_2 \Bigr]
\nn\\&&\hskip0.8cm\null
+ 8 \Bigl[ - {\rm Li}_3(1-z) + \frac{7}{3} \log^3 z
  + \frac{1}{2} \log^2 z \, \log(z-1)
\nn\\&&\hskip1.4cm\null
  - \frac{1}{2} \log z \, \log^2(z-1)
  - 17 \zeta_2 \log z \Bigr] \biggr\}
\nn\\&&\hskip0.0cm\null
- 4 \pi^2 \biggl\{ - \frac{1}{\e^2} + \frac{6\log z}{\e}
- 14 (\log^2 z - \zeta_2 )
\nn\\&&\hskip1.4cm\null
- 2 \log z \, \log(z-1) \biggr\} \,.  \label{discA1D1omzzgt1}
\eea
Finally, the $A_1$ discontinuity of $D_2(1-z)$ is
\bea
\mathrm{disc}_{A_1} D_2(1-z)
&=&  2 \im \pi \biggl\{
\frac{\log|1-z|^2}{\e^2}
- \frac{2}{\e} \log|z|^2 \, \log|1-z|^2
+ 4 {\rm Li}_3(z) + 4 {\rm Li}_3\Big( \frac{-z}{1-z} \Big)
\nn\\&&\hskip0.8cm\null
- 2 \log(1-z) \Big[ {\rm Li}_2(z) - {\rm Li}_2(\zb)
                 - \log^2(1-\zb) - 6 \zeta_2 \Bigr]
\nn\\&&\hskip0.8cm\null
- \frac{1}{3} \log^3|1-z|^2
+ 2 \log^2|z|^2 \, \log|1-z|^2
- 22 \zeta_2 \log|1-z|^2 \biggr\}
\nn\\&&\hskip0.0cm\null
- 4 \pi^2 \biggl\{ \frac{\log|1-z|^2}{\e}
       - 2 \log|z|^2 \, \log|1-z|^2 \biggr\} \,.
\label{discA1D2omz}\\
&=& \frac{1}{2} \mathrm{disc}_{A_1} D_1(z)
- 4 \pi^2 \biggl\{ \frac{\log|1-z|^2}{\e}
       - 2 \log|z|^2 \, \log|1-z|^2 \biggr\}
\nn\\&&\hskip0.0cm\null
+ 2 \im \pi \biggl\{
\frac{\log|1-z|^2}{\e^2}
- \frac{2}{\e} \log|z|^2 \, \log|1-z|^2
- \frac{1}{12} \log^3|1-z|^2
\nn\\&&\hskip1.2cm\null
+ 2 \log^2|z|^2 \, \log|1-z|^2
- 16 \zeta_2 \, \log|1-z|^2
\nn\\&&\hskip1.2cm\null
+ \frac{1}{4} \log \left(\frac{1-z}{1-\bar z}\right)
  \Bigl[  \log^2 \left( \frac{1-z}{1-\bar z} \right)+ 4 \pi^2 \Bigr] \biggr\} \,.
\label{discA1D2omzALT}
\eea
The second form (\ref{discA1D2omzALT}) shows that the linear combination
$D_2(1-z) - \tfrac{1}{2} D_1(z)$
has a discontinuity that is purely logarithmic.  It
illustrates more simply the unambiguous behavior
as $z$ approaches the real axis with ${\rm Re}(z) > 1$.
The last term in \eqn{discA1D2omzALT} is the only one with
a potential ambiguity, but it cancels because
$\log\left(\frac{1-z}{1-\bar z}\right) \to 2\im\pi$,
and $(2\im\pi)^2 + 4\pi^2 = 0$.
In this limit, $D_2(1-z)$ becomes
\bea
\mathrm{disc}_{A_1} D_2(1-z)\bigl|_{z>1,\ {\rm real}}
&=& 2 \im \pi \biggl\{
\frac{2 \log(z-1)}{\e^2} - \frac{8 \log z \log(z-1)}{\e}
\nn\\&&\hskip0.8cm\null
- 4 ( {\rm Li}_3(1-z) - \zeta_3 )
- 2 \log z \Bigl[ \log^2(z-1) + \pi^2 \Bigr]
\nn\\&&\hskip0.8cm\null
+ 4 \Bigl[ 4 \log^2 z - 7 \zeta_2 \Bigr] \log(z-1) \biggr\}
\nn\\&&\hskip0.0cm\null
- 4 \pi^2 \biggl\{ \frac{2 \log(z-1)}{\e} - 8 \log z \, \log(z-1)
\biggr\} \,.  \label{discA1D2omzzgt1}
\eea

To summarize, we combined the tripole contributions to the two-loop
soft factor into four sets of SVHPLs : $\{D_i(z), D_{i}(1-z)\},\ i=\{1,2\}$,
given in eqs.~(\ref{D1value})--(\ref{D2omzvalue}).
Each of the $D_i$'s can be defined in either region $A_0$, or 
in $A_1$  via analytic continuation.  The other regions are not
needed explicitly, due to the freedom in labeling the unordered
tripole $\{i,j,k\}$.
After analytic continuation, the $D_i$ are not single-valued functions,
in the sense that they cannot be written in terms of the
$\cL_{\vec{w}}$ functions, but they are unambiguously defined
in the physical domain where $\bar{z}$ is the complex conjugate of $z$;
that is, where $\Delta = z-\bar{z}$ is purely imaginary,
{\it i.e.}~$1- 2u -2 v +(u-v)^2 \leq 0$.
We emphasize that this is a nontrivial property, which is not satisfied
by the discontinuities of arbitrary linear combinations of SVHPLs.

For negative-helicity emission, one should swap $z \leftrightarrow \bar{z}$
everywhere, as well as letting $\ab{ab} \leftrightarrow \sqb{ba}$ in the
rational prefactors, but the manifest `$\im$'s in the discontinuity formulas
should {\it not} be flipped.
The results for the $D_i$ functions in region $A_0$, their symbols,
and their discontinuities for the analytic continuation into
the $A_1$ region, are all included in an ancillary file
to this arXiv submission. 

%================================================
\vspace{-0pt}
\section{Applications}
\label{sec:applications}\vspace{-4pt}
%================================================

%================================================
\vspace{-0pt}
\subsection{Soft limits of scattering amplitudes}
\label{subsec:soft_lim_amps}\vspace{-4pt}
%================================================

The two-loop soft factor predicts the soft-gluon limit of a generic
two-loop $n$-point scattering amplitude, when the corresponding
$(n{-}1)$-point amplitude without the soft gluon is known.
Such limits can provide stringent checks of two-loop
amplitudes, starting at five points. The tripole terms contribute to amplitudes
beyond the planar limit, which are only just beginning to become available
at five points.

As an example, we can construct the soft limit of the full color
two-loop five-gluon maximally helicity violating~(MHV) amplitude in
$\NeqFour$~\cite{Abreu:2018aqd,Chicherin:2018yne},
using the well-known full color two-loop four-gluon MHV
amplitude~\cite{Bern:1997nh,SmirnovDoubleBox,Tausk:NPDoubleBox},
\bea
 && \lim_{q^\mu \to 0} \big| \mathcal{M}_5^{{\rm MHV},(2)}(1^-,2^-,3^+,4^+,q^+)
                   \big\rangle  \nn\\
 &=& \sum_{\ell=0}^2\! \bom{S}^{+,(\ell)}(\{\beta_m\})
    \big| \mathcal{M}_4^{{\rm MHV},(2-\ell)}(1^-,2^-,3^+,4^+) \big\rangle \,.
  \label{eq:21}
\eea
The $1/\e$ infrared singularities on both sides of \eqn{eq:21}
agree with those obtained from the so-called 
dipole formula~\cite{Catani:1998bh,Sterman:2002qn,Aybat:2006mz,Dixon:2008gr,Becher:2009cu,Gardi:2009qi}, which provides a strong check of our result.
Furthermore, at ${\cal O}(\e^0)$, we have evaluated both sides
explicitly in the trace basis for five external gluons,
and the symbol~\cite{Goncharov:2010jf}
of our soft-emission formula matches perfectly the soft limit of the
symbol-level results for the two-loop five-particle scattering amplitudes in
$\mathcal{N}=4$ SYM \cite{Abreu:2018aqd,Chicherin:2018yne}.  
We anticipate that the knowledge of the soft factor provided
here at function level will play an important role in fixing
certain beyond-the-symbol constants in the supersymmetric amplitude,
as well as for amplitudes with less supersymmetry in the future.

%================================================
\vspace{-0pt}
\subsection{Soft-collinear limit}
\label{subsec:soft_collinear}\vspace{-4pt}
%================================================

As a further interesting application of our result, we can derive the
soft-gluon limit of the two-loop collinear splitting amplitudes for $g\to gg$, 
$q \to qg$, or $\bar{q} \to \bar{q}g$
from the two-loop soft factor computed above.
In the case of timelike splitting, where an off-shell parton splits into
two on-shell partons in the final state, the two-loop splitting amplitudes
have been computed by
several groups~\cite{Bern:2004cz,Badger:2004uk,Duhr:2013msa,Duhr:2014nda}. 

However, for splitting amplitudes with spacelike kinematics, two-loop
splitting amplitudes are so far unknown except for
their singular pieces~\cite{Catani:2011st}. Before we discuss the soft limit
for both timelike and spacelike splitting at two loops, let us briefly comment
on the relevant collinear kinematic setup, establish our conventions,
and discuss the collinear limit of the tree-level and one-loop
soft factors. 

To be specific, consider an $(n+1)$-point scattering process, where $q$ 
is the momentum of an \emph{outgoing} soft gluon
which is collinear to $p_1$.  We distinguish two cases:
\vspace{-0.4cm}
\begin{itemize}
 \item timelike splitting:  	particle 1 is an \emph{outgoing} parton with momentum $p_1$,
 \item spacelike splitting: 	particle 1 is an \emph{incoming} parton with momentum $-p_1$ \newline 
 	\phantom{spacelike splitting:}				(in all outgoing conventions).
\end{itemize}
Let us define the longitudinal momentum fraction $x_q$ carried by 
the soft gluon with respect to the parent parton $P$, such that 
\begin{align} 
q \rightarrow  x_q \, P \,,  \quad    p_1 \rightarrow (1-x_q)\, P\, ,    \qquad  \text{where} \quad 
P^\mu  \equiv  q^\mu + p_1^\mu   - \frac{ (q \cdot p_1) \, n^{\mu }}{  (q + p_1) \cdot n  }  
\end{align}
is a lightlike momentum, $P^2=0$, and $n^\mu$ defines an auxiliary lightlike vector. 
In the case where $q$ is soft,  $P \cong p_1$, so that $|x_q| \ll 1$. Depending on whether we consider timelike or spacelike 
splitting, the sign of $x_q$ changes, from $x_q>0$ to $x_q<0$, respectively.

\medskip 

Before discussing the loop-level analysis of the soft-collinear limit, let us consider the collinear 
limit of the tree-level soft-factor $\bom{S}^{+,(0)}_a$ in eq.~(\ref{eq:3}),
\begin{align}
\bom{S}^{+,(0)}_a = \frac{1}{2n} \sum_{i\neq j} \left(\bom{T}^{a_i}_i - \bom{T}^{a_j}_j \right) \frac{\ab{ij}}{\ab{iq}\ab{qj}} 
			    = \frac{1}{n} \sum_{j<i} \left(\bom{T}^{a_i}_i - \bom{T}^{a_j}_j \right) \frac{\ab{ij}}{\ab{iq}\ab{qj}}\,.
\end{align}
We would like to consider the (timelike) collinear limit $q\parallel p_1$
(or $q\parallel 1$ for short). The only terms containing the singular
factor $\ab{q1}$ are those with $j=1$, 
\begin{align}
\label{eq:soft_coll_tree}
\lim_{q\parallel p_1} \left[ \bom{S}^{+,(0)}_a \right] =  \frac{1}{n} \sum^n_{i =2} \left(\bom{T}^{a_i}_i - \bom{T}^{a_1}_1 \right) \frac{\ab{i1}}{\ab{iq}\ab{q1}}\,.
\end{align}
In the collinear limit (with $q$ also soft), we can now use,
\begin{align}
 \lam{1} \simeq \sqrt{1{-}x_q}\, \lam{P} \simeq \lam{P}, \quad \lam{q} \simeq \sqrt{x_q} \,\lam{P} 
 \quad \Rightarrow \quad \frac{\ab{i1}}{\ab{iq}\ab{q1}} \simeq  \frac{1}{\sqrt{x_q}\, \ab{q1}}\,,
\end{align}
which is independent of $i$, so that we can use color conservation 
$\sum^n_{i=2} \bom{T}^{a_i}_i = - \bom{T}^{a_1}_1$ in the first term 
of (\ref{eq:soft_coll_tree}). In the second term, we get $(n-1)$ times 
the same contribution, which combined with the first term cancels the 
factor of $1/n$. We finally obtain the $q\parallel 1$ collinear limit of the tree-level 
soft factor,
\begin{align}
\label{eq:soft_coll_splitting_tree}
\lim_{q\parallel p_1} \left[ \bom{S}^{+,(0)}_a \right] =   \bom{T}^{a_1}_1 \frac{1}{\sqrt{x_q}\, \ab{1q}} \equiv \textbf{Sp}^{(0)}\,.
\end{align}
Note that the full collinear splitting function depends on the
helicities of the collinear partons, see {\it e.g.}~ref.~\cite{Bern:2004cz}
(there it is denoted by $\textbf{Split}^{(0)}_{-\lambda}$);
the helicity of $P$ is indicated by the helicity label $\lambda$. 
(For a summary of tree-level helicity splitting amplitudes,
see {\it e.g.}~ref.~\cite{Bern:1994zx}
or Appendix A of ref.~\cite{Bern:1999ry}.)
Here, however, we are considering the soft limit of a positive helicity gluon, 
and the helicity information of the collinear splitting drops out,
in the sense that the splitting amplitudes either vanish
(due to helicity conservation rules) or become the same function.
Thus we drop the helicity label on $\textbf{Sp}$ in the following.
The same discussion is also applicable for spacelike splittings,
which can be obtained easily by crossing at tree level,
replacing $x_q \to -x_q$ in \eqn{eq:soft_coll_splitting_tree}.

\medskip

Next we consider the spacelike collinear limit of the one-loop
soft factor~(\ref{eq:4}),
 \begin{align}
 \label{eq:oneloopsp1}  
 \hspace{-.5cm}
 \lim_{q \parallel p_1} 
 \left[ \bom{S}^{+,(1)}_{a} \right] = &\
 \sum_{k \neq 1} \bom{T}^{a_1}_1  \frac{1}{\sqrt{-x_q}\, \ab{1q} }  \,  \left( \frac{\mu^2}{x_q s_{1q} } \right)^{\e} \,   f_{a a_1 a_k} \, \bom{T}^{a_k}_k  \,  
  \exp \big[ (-1)^{\lambda_{kq}+1}  \im \pi \e \big] \,   C_1(\e)  \,.
 \hspace{-.5cm}
 \end{align} 
Note that $x_q s_{1q}$ is always positive. 
Following the definition of the color generators in the adjoint representation, we write $ (T_q^a)_{bc}  \equiv  - f^{abc}$.  
The sum over dipoles give a contribution to the one-loop splitting amplitude, 
\begin{align}
  \label{eq:oneloopsp2}
  \bom{\rm Sp}^{(1)} \quad 
  \hskip -.3cm
  \stackrel{q-\text{soft}}{\simeq}
  \hskip 0cm
   -  \left( \frac{\mu^2}{x_q s_{1q} } \right)^{\e} C_1(\e) 
    \sum_{k \neq 1}  \bom{T}_q \mcdot \bom{T}_k \exp \big[ (-1)^{\lambda_{kq}+1}  \im \pi \e \big]\, \bom{\rm Sp}^{(0)} \,,
\end{align}
where $\bom{\rm Sp}^{(0)} = \bom{T}_1  \frac{1}{\sqrt{-x_q}\,\ab{1q}}$ is the soft limit of the spacelike tree-level splitting amplitude, which is obtained from the timelike splitting amplitude of eq.~(\ref{eq:soft_coll_splitting_tree}) by crossing (including replacing $x_q\to-x_q$).

Note that the one-loop splitting amplitude in eq.~\eqref{eq:oneloopsp2} already violates 
the so-called \emph{strict collinear factorization}, which states that the splitting amplitude 
should only depend on the quantum numbers~(color, spin and kinematics) of the parent parton 
and of the splitting pair~\cite{Catani:2011st} and not on any information of the non-collinear partons. 
This condition is violated at the amplitude level due to the existence of the $\lambda_{kq}$-dependent 
phase factor in eq.~\eqref{eq:oneloopsp2}, which obstructs the use of color conservation 
that was used in the tree-level case. Fortunately, the $\lambda_{kq}$ dependence is purely imaginary 
and cancels out for cross-section level quantities at this perturbative order, which involve only the combination 
$\bom{\rm Sp}^{(0)} \bom{\rm Sp}^{(1)\dagger} + \bom{\rm Sp}^{(0)\dagger} \bom{\rm Sp}^{(1)}$. See the discussion around eq.~(\ref{eq:dipole_soft_factor_re_im_manifest}) for more details.

In the timelike case, the factor $(V_{1k}^q)^\e$, which led to the $k$-dependent phase in the spacelike case in \eqn{eq:oneloopsp2}, behaves differently: The $k$-dependent phases cancel between the numerator and denominator of $(V_{1k}^q)^\e$ in the timelike case.

%================================================
\vspace{-0pt}
\subsubsection*{Soft-collinear analysis for timelike splitting at two loops}
\label{subsubsec:soft_collinear_timelike}\vspace{-4pt}
%================================================
%
For timelike splitting where the outgoing soft gluon with momentum $q$
is collinear to the \emph{outgoing} particle 1 with momentum $p_1$,
we have $\sab{1q}>0$ and $x_q>0$. The soft limit of the two-loop splitting
amplitude can be obtained by taking the collinear limit of the two-loop
dipole and tripole contributions,
using eqs.~(\ref{eq:6}) and (\ref{tripolesumALT}),
\begin{align}
 \label{eq:348}  
 \lim_{q \parallel p_1}  \, \left[ \bom{S}^{+,(2)}_a  \right]=
   \lim_{q \parallel p_1}  \,  \left[   
    \sum_{j \neq 1 } \bom{S}^{+,(2)}_{a, 1j }  
  - \frac{1}{4} \sum_{ \{ 1 , j, k \} }  \bom{S}^{+,(2)}_{a, \{ 1, j,k  \} }  
  \right] \,.
\end{align}
The collinear limit of the dipole soft-emission term at two loops is straightforward; it is entirely captured by the collinear limit of $\left(V^q_{1j}\right)^2$ and the tree-level eikonal factor in eq.~(\ref{eq:7}) and therefore we do not discuss it any further for the timelike case.

%%%%%%%%%%%%%%%%%%%%%%
\begin{figure}[ht!]
    \centering
    \begin{subfigure}[c]{0.1\textwidth}
      \raisebox{0pt}{\includegraphics[scale=0.5]
         {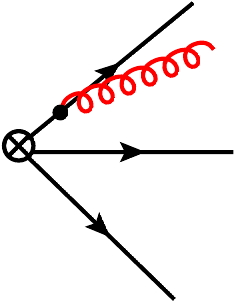}}
        \caption{}
        \label{fig:ra}
    \end{subfigure}
    \qquad \qquad
    \begin{subfigure}[c]{0.17\textwidth}
      \includegraphics[scale=0.5]{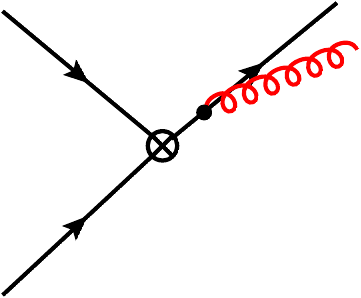}
        \caption{}
        \label{fig:rb}
    \end{subfigure}
   \qquad
    \begin{subfigure}[c]{0.1\textwidth}
      \includegraphics[scale=0.5]{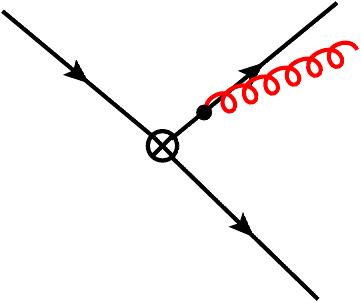}
        \caption{}
        \label{fig:rc}
    \end{subfigure}
\\
    \begin{subfigure}[c]{0.1\textwidth}
      \includegraphics[scale=0.5]{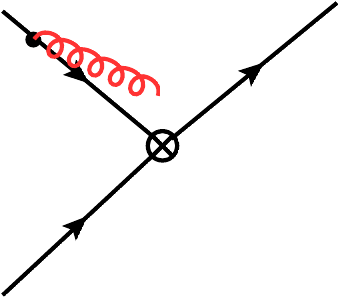}
        \caption{}
        \label{fig:rd}
    \end{subfigure}
\qquad\qquad\qquad
    \begin{subfigure}[c]{0.1\textwidth}
      \includegraphics[scale=0.5]{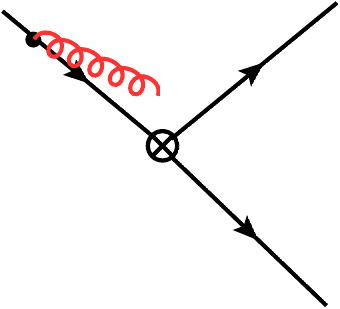}
        \caption{}
        \label{fig:re}
    \end{subfigure}
    \caption{Kinematics of a tripole contribution to the soft factor in the
      collinear limit in different regions: (a) timelike splitting in $A_0$,
      (b) timelike splitting in $A_1$, $A_2$, $A_3$,
      (c) timelike splitting in $A_4$, $A_5$, $A_6$,
      (d) spacelike splitting in $A_1$, $A_2$, $A_3$,
      (e) spacelike splitting in $A_4$, $A_5$, $A_6$.
      The curly gluon line represents the outgoing soft gluon with momentum $q$,
      while the lines with arrows are the tripole legs $i,j,k$.
      The arrow of time flows from left to right.}
\label{fig:region}
\end{figure}

Due to the nontrivial kinematic dependence, the collinear limit of the tripole soft-emission term given in eq.~(\ref{eq:tripolesum}) is more interesting. For timelike splitting kinematics, and assuming at most two partons in the initial state, there are 
three cases to consider (for an outgoing soft gluon), c.f. figs.~\ref{fig:ra}, \ref{fig:rb}, \ref{fig:rc}: 
\newline
a) The collinear limit of the tripole term in the $A_0$ region, where all hard partons $\{i,j,k\}$ are 
outgoing and the soft gluon is collinear to either $i,\,j$ or $k$, see fig.~\ref{fig:ra}. 
\newline
b) The collinear limit of a tripole with two incoming hard partons and one outgoing hard parton to which the soft gluon is collinear (regions $A_{1,2,3}$, fig.~\ref{fig:rb}). 
\newline
c) The collinear limit of a tripole with one incoming and three outgoing lines (regions $A_{4,5,6}$, fig.~\ref{fig:rc}) 
where the outgoing soft gluon is collinear to an outgoing hard parton. 

We first discuss the timelike collinear limit of the soft-emission tripole of eq.~(\ref{eq:tripolesum}) in the $A_0$ region of table~\ref{tab:1}. In principle, there are three possibilities to take the collinear limit, where the soft gluon $q$ becomes collinear to either of the three hard lines $\{i,j,k\}$ to be identified with $p_1$. Since the tripole formula (\ref{eq:tripolesum}) is completely symmetric in $\{i,j,k\}$ it suffices to look at a single collinear limit, which we take to be $i=1$ for convenience. Any other implementation of the collinear limit, {\it e.g.}~with $j=1$ or $k=1$, can then be used as a cross check of our setup. In the $q \parallel  p_1 ({=}p_i)$ collinear limit, only the first eikonal factor in eq.~(\ref{eq:tripolesum}) (and not the $\{i\lra j\}$ exchange term) contains the singular denominator $1/\ab{iq}$, so that we need to analyze the limit of the transcendental functions $D_1(z,\zb)$ and $D_{2}(z,\zb)$ in eq.~(\ref{eq:tripolesum}) as $z = z^{ij}_k = \ab{kj}\ab{iq}/(\ab{ij}\ab{kq})\to 0$ (and likewise $\zb \to 0$).  The only SVHPLs that do not vanish at the origin are
$\cL_{0^n}$, $n=1,2,\ldots$.  Since these functions don't appear alone in
any term in \eqns{D1value}{D2value} for $D_1$ and $D_2$, we find that
\begin{align}
  \lim_{z,\zb\to 0} D_1(z,\zb) = \lim_{z,\zb\to 0} D_2(z,\zb) = 0\,.
  \label{Divanishat0}
\end{align}
This result can also be seen from the explicit representations of these
functions in terms of classical polylogarithms in eqs.~(\ref{D1polylog})
and (\ref{D2polylog}), respectively.  Thus the $D_i(z,\zb)$ provide a power
suppression that cancels the leading $1/\ab{iq}$ singularity in the eikonal
factor in \eqn{eq:tripolesum}.

Due to the manifest tripole symmetry under the $i\leftrightarrow j$ exchange, the $q\parallel j$ collinear limit ($z \to 1$) of the tripole formula is equally suppressed. A little more nontrivially, for consistency of the tripole symmetry, one can also analyze the $q\parallel k$ collinear limit of (\ref{eq:tripolesum}). Here both eikonal factors have the $1/\ab{qk}$ pole, and we wish to send $z,\bar{z} \to \infty$.  However, the kinematic factors become the same in both terms ($\lam{q} \to \sqrt{x_q}\, \lam{k} $, $\lamt{q} \to \sqrt{x_q}\, \lamt{k} $):
\bea
\frac{\ab{ik}}{\ab{iq}\ab{qk}} \left(V^q_{ik}\right)^2
&\stackrel{q\parallel k}{\longrightarrow}&
\frac{1}{\sqrt{x_q}\, \ab{kq}} \left[\frac{\mu^2}{x_q (-\sab{kq})}\right]^{2\e}
\,, \label{Viklim}\\
\frac{\ab{jk}}{\ab{jq}\ab{qk}} \left(V^q_{jk}\right)^2
&\stackrel{q\parallel k}{\longrightarrow}&
\frac{1}{\sqrt{x_q}\, \ab{kq}} \left[\frac{\mu^2}{x_q (-\sab{kq})}\right]^{2\e}
\,.
\label{Vjklim}
\eea
Also, the color factors associated with $D_1(z)$ and $D_2(1-z)$, as well as with $D_2(z)$ and $D_1(1-z)$, are the same up to a sign.  Finally, we find that 
the $D_i$ functions all have the same logarithmic divergence as $z \to \infty$, 
\begin{align}
\lim_{z\to\infty} D_1(z)=\lim_{z\to\infty} D_2(z)=\lim_{z\to\infty} D_1(1-z)=\lim_{z\to\infty} D_2(1-z)\,.
\end{align}
Therefore the $q\parallel k$ collinear limit is also power suppressed as expected. We conclude that in the $A_0$ region, the tripole emission contribution is subleading by a power of $q$ when taking the collinear limit inside the soft limit.

The second case to consider is the tripole with two incoming hard partons and one outgoing hard parton to which the soft gluon is collinear, fig.~\ref{fig:rb}. 
This collinear limit requires the analysis of the discontinuity of the tripole term to the $A_1$ region where we have judiciously chosen the 
collinear hard parton to be labelled by $i=1$. Looking at the discontinuities of the $D$ functions associated with the eikonal factor that contains 
the $\ab{iq}$ pole, one can indeed check from eqs.~(\ref{discA1D1}) and (\ref{discA1D2}) that these also vanish in the collinear limit $z\to0$,
\begin{align}
\lim_{z,\zb \to 0} \text{disc}_{A_1} D_1(z,\zb) = \lim_{z,\zb \to 0} \text{disc}_{A_1} D_2(z,\zb) = 0\,.
\end{align}
Finally, since the configurations with two 
outgoing hard partons (regions $A_4,\, A_5$ and $A_6$ in table \ref{tab:1}, c.f. fig.~\ref{fig:rc}) do not require an
analytic continuation, the timelike collinear limits of those tripoles are also suppressed by a power of the soft momentum $q$. 

In conclusion, for timelike splitting kinematics, all tripole collinear limits are power suppressed 
and the leading collinear singular information is provided by the dipole emission contribution.  (In the timelike case, the two-loop
dipole contribution is also rather innocuous,
because there is no $k$-dependent phase, as was mentioned at the end of
the previous subsection for the one-loop case.)
The subleading nature of tripole emission in the soft-collinear limit
is quite reasonable, because the soft-collinear limit of a tripole is
conformally equivalent to making the other two hard legs collinear.
For example, $q\parallel p_i$ and $p_j \parallel p_k$ both send $z_k^{ij} \to 0$.
When the two hard partons $p_j$ and $p_k$ are collinear and both are incoming 
or both outgoing, the soft gluon cannot resolve the independent colors
of the two particles as a consequence of color coherence; it just sees the sum,
and therefore the emission is dipole-like. However, in the spacelike case
where $p_j$ and $p_k$ are incoming and outgoing, respectively, this simple
picture breaks down. The physical origin of the breakdown of color coherence
in this case is related to the Feynman $\im \varepsilon$ prescription,
and therefore to the causality of the theory. Ultimately, this can potentially lead to
a violation of strict factorization, but as we shall discuss in the next
subsection these terms cancel at the level of the cross section.

%================================================
\vspace{-0pt}
\subsubsection*{Soft-collinear analysis for spacelike splitting at two loops and strict collinear factorization violation}
\label{subsubsec:soft_collinear_spacelike}\vspace{-4pt}
%================================================

Let us now discuss spacelike splitting at two loops, where particle 1
is an \emph{incoming} parton with momentum $-p_1$
(in all outgoing conventions) so that $\sab{1q}<0$, 
and the longitudinal momentum fraction is $x_q<0$.
First we relabel indices in the two-loop soft-collinear
factor~(\ref{eq:348}) to obtain
\begin{align}
 \label{eq:348p}  
 \lim_{q \parallel p_1}  \,  \bom{S}^{+,(2)}_a =
   \lim_{q \parallel p_1}  \left[  \sum_{k \neq 1 } \bom{S}^{+,(2)}_{a, 1k }  
     - \frac{1}{4} \sum_{ \{ i , 1, k \} }  \bom{S}^{+,(2)}_{a, \{ i, 1,k  \} }   \right]
   \,.
\end{align}
This labeling will allow us to utilize the
$A_1$ discontinuity discussed in section \ref{subsec:analytic_cont}.

The analysis of the collinear limit of the dipole term in eq.~(\ref{eq:348p})
follows closely the one-loop dipole analysis, eq.~\eqref{eq:oneloopsp1}. 
The sum over dipoles gives a contribution to the two-loop collinear
splitting amplitude, 
\begin{align}
  \label{eq:22}  
  \bom{\rm Sp}^{(2)}\Big|_{\rm dipole}
  \stackrel{q-\text{soft}}{\simeq}
   -  \left( \frac{\mu^2}{x_q s_{1q} } \right)^{2\e} C_2(\e) 
    \sum_{k \neq 1}  \bom{T}_q \mcdot \bom{T}_k \exp \big[ (-1)^{\lambda_{kq}+1}  2\im \pi \e \big] \bom{\rm Sp}^{(0)} \,.
\end{align}
For $L= 1 \;\text{or}\; 2 $ we can rewrite the Hermitian
part of the dipole terms in eqs.~(\ref{eq:oneloopsp1}) and (\ref{eq:22})
by applying color conservation: 
\begin{align}
  \sum_{k \neq 1 } \bom{T}_q \cdot \bom{T}_k
  = - \bom{T}_q \cdot (\bom{T}_1 +\bom{T}_q)
  = - \frac12  (C_{1q} - C_1 )  - \frac12 C_q
  = - \frac12 C_A  \,, 
\end{align}
where $C_{1q}$, $C_1(=C_{1q})$, and $C_q = C_A$
are the Casimir coefficients of parton $P$,
parton 1, and the soft-collinear gluon, respectively.
The result is
\bea
\bom{\rm Sp}^{(L)} \Big|_{{\rm dipole}} &\stackrel{q-\rm soft}{\simeq}&
 \left( \frac{\mu^2}{x_q s_{1q} } \right)^{L\e} 
 \!  C_L (\e)
  \bigg\{
 \im \sin (L \pi \e ) 
    \sum_{k \neq 1}  (-1)^{\lambda_{kq}}  \bom{T}_q \mcdot \bom{T}_k 
    + \frac{C_A}{2}  \cos (L\pi  \e) \bigg\}  \bom{\rm Sp}^{(0)} \,.
\nonumber\\
&~&~~~ \label{eq:dipole_soft_factor_re_im_manifest}
\eea
The factorization-violating kinematic dependence of
the $L$-loop dipole term is anti-Hermitian.
The Hermitian part proportional to $\cos(L\pi\e)$ strictly factorizes. 

We turn now to the tripole emission contribution.  As mentioned earlier,
it vanishes in all collinear limits for timelike splittings.
However, for spacelike splitting kinematics it is quite nontrivial.
Eq.~\eqref{eq:348p} contains a sum over two types of tripoles, 
corresponding to the two different kinematic configurations 
where $\lambda_{ik} =1$ and $\lambda_{ik} =0$. 
In the first situation, $\{ i,k, q \}$ are outgoing (fig.~\ref{fig:re});
thus, according to table \ref{tab:1}, we consider 
the analytic continuation of the tripole term to the $A_5$ region.
Since we set $j=1$, the tripole contributions in the collinear limit
come from the $i \leftrightarrow j$ terms in \eqn{eq:tripolesum}, 
and from the region where $z \equiv z_{k}^{i1} \rightarrow 1$.
Because the $A_5$ region is equivalent to the $A_0$ region, we find  
\begin{align}
\label{tripoleA5}
 \lim_{z, \bar{z}  \rightarrow 1 } \left[ \bom{S}^{+,(2)}_{a, \{ i,1, k\}} \Big|_{A_5}\right]
 = &\    \bom{T}^{a_1}_1  \frac{1}{\sqrt{-x_q} \ab{1q} }   \,  
 \left( \frac{\mu^2}{x_q s_{1q} } \right)^{2\e}
      \xi_\e \, {\rm{exp}} \left[ 2 \im \pi \e \right]
      \,  2 \, \bom{T}^{a_i}_i\bom{T}^{a_k}_k \,  
 \times  
  \\
&  \lim_{z, \bar{z}  \rightarrow 1 } \,  
 \Bigl[ f^{a  a_i  b } f^{b a_1 a_k} \,  D_1( 1- z, 1- \bar{z} )	 + f^{a a_1 b} f^{b a_k a_i}   \, D_2(1-z, 1- \bar{z}) \Bigr] \nn \,, 
\end{align} 
with (see \eqn{Divanishat0})
\begin{align} 
  \lim_{z, \bar{z}  \rightarrow 1} D_1 (1-z, 1- \bar{z} )  = 0,
  \quad  \lim_{z, \bar{z}  \rightarrow 1}  D_2 (1-z, 1- \bar{z} )  = 0\, . 
\end{align}
We conclude that for $\lambda_{ik} = 1$,  $ \bom{S}^{+,(2)}_{a, \{ i,1, k\}} $ vanishes in the collinear limit.

For the second type of tripole with $\lambda_{ik} =0$, either $p_i$ or $p_k$ is incoming, c.f. fig.~\ref{fig:rd}. 
Without loss of generality, we label the incoming leg by $k$ and 
consider the analytic continuation of the tripole term to the $A_1$ region 
where $ \{ 1, k \} $ are incoming and $\{ i,q \} $ are outgoing. 
In this region, the collinear limit of $\bom{S}^{+,(2)}_{a, \{i,1,k\}} $ 
is obtained from the collinear limit of its $A_1$ discontinuity. 
\begin{align}
\label{eq:354}
 \lim_{z, \bar{z}  \rightarrow 1} \left[\bom{S}^{+,(2)}_{a, \{i,1, k\}}\Big|_{A_1}  \right]
 =  &\ \lim_{z, \bar{z}  \rightarrow 1}  \,
 \text{disc}_{A_1} \, \bom{S}^{+,(2)}_{a, \{i,1, k\}}  \nn \\
 = &\ \bom{T}^{a_1}_1  \frac{1}{\sqrt{-x_{q}}\, \ab{1q} }  \,  
\left( \frac{\mu^2}{x_q s_{1q} } \right)^{2\e}
(1+\e^2 \zeta_2) \,
  {\rm{exp}} \left[- 2 \im \pi \e \right]   \nn \\
 & \hskip0.2cm \times\
  2 \, \bom{T}^{a_i}_i\bom{T}^{a_k}_k \,  
    \lim_{z, \bar{z}  \rightarrow 1} \,  
 \Bigl[ f^{a  a_i  b } f^{b a_1 a_k} \,  {\rm{disc}}_{A_1} D_1( 1- z, 1- \bar{z} )	 \nn \\
&   \hskip3.2cm    
       + f^{a a_1 b} f^{b a_k a_i}   \, {\rm{disc}}_{A_1} D_2(1-z, 1- \bar{z}) \Bigr]\,, 
\end{align} 
where 
\begin{align}
 & \lim_{z, \bar{z}  \rightarrow 1} {\rm{disc}}_{A_1} D_1( 1- z, 1- \bar{z} ) =   4\pi^2 \, \left( \frac{1}{\epsilon^2 }  +\frac{ 2  \im \pi  }{ \epsilon} - 14 \, \zeta_2  \right) \, , \\
& \lim_{z, \bar{z}  \rightarrow 1}   {\rm{disc}}_{A_1} D_2( 1- z, 1- \bar{z} ) 
 =   2 \im \pi  \left\{  \frac{\log |1-z|^2 }{ \epsilon^2 }  +  2  \im \pi  \frac{\log |1-z|^2 }{ \epsilon}     \right.  \notag \\ 
  &  \hspace{1.0cm}  \left.   +   \frac{1}{3} \log \left(\frac{1-z}{1- \bar{z}}\right) \left[   \log^2 \left(\frac{1-z}{1- \bar z}\right) + 4 \pi^2 \right] + 
  4 \zeta_3 - 14 \, \zeta_2  \log |1-z|^2   \right\} \,,
\end{align}
and
\bea
\log(1-z) = \log(1-z_k^{i1}) = \log z_k^{1i} \,, \qquad
\log(1-\bar z) = \log \bar{z}_k^{1i} \,.
\label{argumentzk1i}
\eea
Applying the color Jacobi identity, we decompose the color factors
in \eqn{eq:354} into structures that are even and odd under the
$i \leftrightarrow  k$ interchange, 
\begin{align}
  f^{a a_i b} f^{b a_1 a_k}  & = - \frac{1}{2 }  \left( T_q^{a_i} T_q^{a_k} + T_q^{a_k} T_q^{a_i}   \right)_{a a_1}  +   \frac12  f^{ b a_k a_i  }   \left( T_q^b \right)_{a a_1} \,,  \\
  f^{a a_1 b} f^{b a_k a_i}   & = -  f^{ b a_k a_i  }   \left( T_q^b \right)_{a a_1}  \,.
\end{align} 
Summing over all external legs, we obtain the two-loop tripole contribution to the splitting amplitude, written in terms of  $1- z_{k}^{i1}  = z_{k}^{1i}$, 
\begin{align}
  \label{eq:23}
  \bom{\rm Sp}^{(2)}\bigg|_{\rm tripole}   
  &\stackrel{q-{\rm soft}}{\simeq} 
   -\frac14 \sum_{\substack{\text{tripoles} \\ \{i,1,k\}}} \bom{S}^{+,(2)}_{a,\{i,1,k\}} \bigg|_{q \parallel  p_1} \nn \\
  &
  =
  \   \left( \frac{\mu^2}{x_q s_{1q} } \right)^{2\e}
   \sum_{ i \neq k \neq 1 } 
  \delta_{0,\lambda_{ik}}  \delta_{1,\lambda_{1k} } 
  	\Bigg\{f^{b a_k a_i }  \bom{T}^b_q\,  \bom{T}^{a_k}_k\,  \bom{T}^{a_i}_i  
 	    \times \bigg[ 			 \\
   &		\hspace{.5cm}
\frac{1}{\e^2} \Big(  \im \pi \log v_k^{1i} {-}  \pi^2\Big)
{-} \frac{ \im\pi^3 }{6}  \log v_k^{1i}
{+} 4  \im \pi \zeta_3
{+} 15 \zeta_4
{+} \frac{8\pi}{3} \big(\mathrm{arg}(z_k^{1i})^3
{-} \pi^2 \mathrm{arg}(z_k^{1i}) \big)
		\bigg] \nn \\
            &
+  \bigg[(\bom{T}_q \mcdot \bom{T}_i ) \, (\bom{T}_q \mcdot \bom{T}_k )
       + (\bom{T}_q \mcdot \bom{T}_k) \, (\bom{T}_q \mcdot \bom{T}_i ) \bigg]
\Big(\frac{\pi^2}{\e^2}
- 15 \zeta_4
\Big) 
 \Bigg\}  \bom{\rm Sp}^{(0)}\,,\nn
\end{align}
where $\mathrm{arg}(z) \equiv \frac{1}{2\im} \log \frac{z}{\zb } \in (-\pi, \pi]$ is the argument of $z$, and the $\delta_{1,\lambda_{1k} }$ is present to enforce our choice that we always label the incoming legs by $\{1,k\}$. Furthermore, we have expanded the phase factor $\exp[-2\im \pi \e]$ in $\e$ and only keep overall terms to order $\mathcal{O}(\e^0)$. As before, the kinematic dependence is given in terms of
\begin{align}
  \label{eq:24}
  v^{1i}_k = \frac{s_{ik} s_{1q}}{s_{1i} s_{kq}} \,, 
  \qquad 
  z^{1i}_k  = \frac{\ab{ki} \ab{1q}}{ \ab{1i} \ab{kq}} \,.
\end{align}
The soft-collinear tripole term $\bom{\rm Sp}^{(2)}\Big|_{\rm tripole}$
in \eqn{eq:23} depends on the color and kinematics of the non-collinear
tripole partons.  It therefore explicitly breaks strict collinear
factorization of scattering amplitudes.
The factorization-breaking $1/\epsilon$ poles in the two-loop splitting
amplitude were given previously \cite{Catani:2011st}.
In the soft limit, they agree with the $1/\epsilon$ poles in
eq.~(\ref{eq:23}).  The two-loop finite terms in eq.~(\ref{eq:23}) 
constitute a new result.  

%======================|Split|^2 ==========

Our results also have implications for factorization violation at the level
of the cross section.  More explicitly, let us consider the perturbative
expansion of the squared splitting probability, $|\textbf{Sp}|^2$.
At leading order, $\cO(\alpha_s)$ and at next-to-leading order,
$\cO(\alpha_s^2)$, strict factorization holds.  At next-to-leading order,
as mentioned earlier, it holds because the $\lambda_{kq}$ dependence is
purely imaginary and cancels from the interference between tree and one-loop,
$\bom{\rm Sp}^{(0)} \bom{\rm Sp}^{(1)\dagger}
+ \bom{\rm Sp}^{(0)\dagger} \bom{\rm Sp}^{(1)}$.
Now let us examine the situation at $\cO(\alpha_s^3)$.
Factorization-breaking contributions 
come from both $ |\bom{\rm Sp}^{(1)} |^2 $ 
and the interference between
$\bom{\rm Sp}^{(2)} $ and $ \bom{\rm Sp}^{(0)}$.   

We first examine the interference of two one-loop splitting amplitudes
in the soft limit, $ |\bom{ \rm Sp}^{(1)}|^2$.  Employing
eq.~(\ref{eq:dipole_soft_factor_re_im_manifest})), we see that it
does not fully factorize, and the non-factorizing part
contains a double pole in $\epsilon$,
 \begin{align} 
\Big|  \bom{\rm Sp}^{(1)}  \Big|^2_{\text{non-fac.}}  \hspace{-0.1cm}
\stackrel{q-\rm soft}{\simeq}   
\hspace{-0.1cm}
 \sum_{ i \neq k \neq 1 }  \! \delta_{0,\lambda_{ik}} \bom{\rm Sp}^{(0)\dagger}
  &	\Bigg\{ 
  \bigg[ - (\bom{T}_q \mcdot \bom{T}_i ) \, (\bom{T}_q \mcdot \bom{T}_k ) 
  - (\bom{T}_q \mcdot \bom{T}_k) \, (\bom{T}_q \mcdot \bom{T}_i ) \bigg] \nn \\
	&\ \times  \left( \frac{\mu^2}{x_q s_{1q} } \right)^{2 \e} 
        \bigg[  \frac{\pi^2 }{\e^2}  - 15\,  \zeta_4  + \cO(\e)  \bigg]   \Bigg\} \bom{\rm Sp}^{(0)}   \,. 
 \end{align} 

Now we consider the interference between $\bom{\rm Sp}^{(2)}$
and $ \bom{\rm Sp}^{(0)}$.  The contributions of the two-loop dipole terms
in $\bom{\rm Sp}^{(2)}$ factorize at the level of the cross section,
for the same reason that strict factorization holds at $\cO(\alpha_s^2)$.
The two-loop tripole term in \eqn{eq:23}
contains both Hermitian and anti-Hermitian parts. 
The latter cancel out in the squared splitting amplitude.  
In the final answer for $|\bom{\rm Sp}|^2$, the factorization-breaking  
$1/\e^2$ pole in $ |\bom{\rm Sp}^{(1)} |^2$ is cancelled by the one in
$\{\bom{\rm Sp}^{(0) \dagger} \bom{\rm Sp}^{(2)}\big|_{\rm tripole} + \rm{h.c.} \}$. 
Combining all remaining terms, we find, 
\begin{multline} 
\label{eq:25}
 \bom{\rm Sp}^{\dagger} \bom{\rm Sp}  \Big|_{\rm non-fac.} \hskip -.5cm   
 \stackrel{ q-\rm soft}{\simeq} 
 \hspace{.05cm}
 \bar{a}^2 g_s^2 \hspace{-.05cm}
 \sum_{ i \neq k \neq 1 }  \! \delta_{0,\lambda_{ik}} \bom{\rm Sp}^{(0)\dagger}
 \Bigg\{ 
  \\
   \hskip .2cm
   2 \pi \im \,  \delta_{1,\lambda_{1k }} \,
   f^{b a_k a_i}  \bom{T}^b_q\,  \bom{T}^{a_k}_k\,  \bom{T}^{a_i}_i 
   \left(  \frac{\mu^2 }{ x_q  s_{1q} }  \right)^{2 \epsilon }
   \bigg[   \Big( \frac{ 1}{\e^2}
-  \zeta_2
                  \Big)   \log v_{k}^{1i} 
 	   +  4 \zeta_3 
		\bigg]  
  \Bigg\}  \bom{\rm Sp}^{(0)} 
+ {\cal O}(\bar{a}^4)  \,.
\hskip -.2cm
\end{multline}
Recall that in our conventions the structure constants $f^{abc}$
are purely imaginary, so in the second line in \eqn{eq:25}
the `$+\rm{h.c.}$' terms result in a projection onto the
kinematic terms containing an explicit `$\im$'.

The color structure in
\eqn{eq:25} can be rewritten as a commutator,
$[ (\bom{T}_q \mcdot \bom{T}_i ) , (\bom{T}_q \mcdot \bom{T}_k ) ]$.
When \eqn{eq:25} is sandwiched between tree amplitudes,
$\langle \mathcal{M}^{(0)}_{n} | \cdots | \mathcal{M}^{(0)}_{n}\rangle$,
and a color sum is performed, the Hermiticity of the operators
$\bom{T}_q \mcdot \bom{T}_i$ allows one to conclude that the color
sum vanishes~\cite{Forshaw:2012bi}.
(A similar cancellation occurs for the $1/\e$ pole with the same color
structure, which appears in two-loop four-point amplitudes~\cite{Bern:2002tk}
but cancels in the color-summed cross section \cite{Glover:2001af}.)

We conclude that for pure QCD splitting processes at order $g\times g^5$,
or ${\cal O}(\alpha_s^3)$,
there is no potential factorization violation visible in the
  soft-collinear limit.
(The failure of strict
factorization at NNNLO for non-inclusive hadron collider processes
has been argued previously, based partly on the structure of $1/\e$ pole
terms associated with Coulomb gluon
exchanges~\cite{Forshaw:2006fk,Forshaw:2008cq,Forshaw:2012bi}.)
An interesting example that can contain such
factorization-violating contributions is the NNNLO QCD corrections to
dijet production at hadron colliders. While the full NNNLO QCD corrections
might still be far away, a shortcut to
investigating potential
factorization-breaking
terms is through the study of precision hadron collider event
shapes~\cite{Gao:2019ojf}, where NNNLO corrections including logarithms
in the event-shape variable are within reach.
We leave the investigation of these important issues to future work.

%================================================
\vspace{-0pt}
\section{Conclusions}
\label{sec:conclusions}\vspace{-8pt}
%================================================

In this paper we computed the exact kinematic and color dependence of
soft-gluon emission in massless gauge theory at the two loop level.
While the dipole terms have a simple kinematic dependence and had been
computed previously~\cite{Li:2013lsa,Duhr:2013msa}, the subleading-color
tripole terms are new, and they depend in an intricate way
on a rescaling-invariant cross ratio.

Using the soft-collinear limit of our results, we could study the soft
limits of two-loop collinear splitting amplitudes for both timelike and
spacelike kinematics.  The timelike behavior was understood
previously~\cite{Bern:2004cz,Badger:2004uk,Duhr:2013msa,Duhr:2014nda}.
In the spacelike case, the infrared singular parts of the two-loop splitting
amplitudes were obtained before in ref.~\cite{Catani:2011st},
with which we find full agreement. Our new results for this case
are the finite contributions, provided in \eqn{eq:23}.
Note that eq.~\eqref{eq:23} is non-zero only when the non-collinear
tripole partons $i$ and $k$ are spacelike separated.  Thus, including
the collinear parton 1, there must be two partons in the initial
state to get a contribution ({\it i.e.}~deep inelastic scattering
does not qualify, while hadronic collisions do).
Both eqs.~\eqref{eq:dipole_soft_factor_re_im_manifest} and
\eqref{eq:23} violate strict collinear
factorization~\cite{Catani:2011st,Forshaw:2012bi},
in the sense that the splitting amplitudes depend on the color and/or
kinematics of some non-collinear hard partons in the process.
For dipole emission, eq.~\eqref{eq:dipole_soft_factor_re_im_manifest},
factorization violation only exists in the imaginary part.
The real part preserves strict collinear factorization, after using
color conservation. Also, the violation in the dipole emission contribution
is independent of kinematics (as a multiple of tree-level splitting).  
On the other hand, the tripole emission, eq.~\eqref{eq:23}, violates
strict collinear factorization with both \emph{color} and \emph{kinematic}
correlations at the amplitude level.

The violation of strict collinear factorization is related to issues
in the breakdown of factorization for transverse-momentum-dependent
PDFs~\cite{Bomhof:2004aw,Collins:2007nk,Vogelsang:2007jk,Rogers:2010dm}, 
double parton scattering~\cite{Gaunt:2018eix,Diehl:2019rdh,Buffing:2017mqm}, 
and the resummation of logarithms in non-global 
observables~\cite{Forshaw:2006fk, Forshaw:2008cq,Balsiger:2018ezi, Becher:2016mmh} 
or event shapes at hadron colliders~\cite{Banfi:2004nk,Stewart:2010pd,Becher:2015gsa,Gao:2019ojf}.
It has been studied in an effective field theory framework which isolates the relevant physical degree of freedom, 
the so-called ``Glauber'', or ``Coulomb'' mode, which is responsible 
for the breaking of collinear factorization~\cite{Rothstein:2016bsq,Schwartz:2017nmr,Schwartz:2018obd}. 
Our explicit result for the two-loop soft factor manifestly shows the breakdown of strict
collinear factorization at the amplitude level in the soft limit. 
However, we have shown the cancellation of potential factorization-breaking terms at the cross-section level as ${\cal O}(\alpha_s^3)$ contributions to single soft-collinear splitting in pure QCD processes 
such as dijet production at hadron colliders. It would be interesting to understand the relation 
between the factorization-breaking terms in the soft factor and the Glauber gluon, by extending 
the one-loop soft current analysis in ref.~\cite{Rothstein:2016bsq} to two loops.

At large $N_c$, the two-loop soft-gluon emission factor has already been
applied to compute the threshold soft corrections to the production of a
color neutral state at hadron colliders at
NNNLO~\cite{Anastasiou:2014vaa,Li:2014afw}.
The results of this paper can be applied as a building block in precision
calculations of generic jet cross sections, including full color dependence.

The analytic behavior of the tripole emission terms for soft-gluon emission at two loops is quite intricate.  The functions $D_1$ and $D_2$ must be well-defined and unambiguous, not only in the Euclidean region $A_0$, but also in other regions obtained by analytic continuation, such as $A_1$.  While any SVHPL is unambiguous in the Euclidean region, it may develop branch cuts in Minkowski regions.  The requirement that these cuts cancel when this region is kinematically accessible, as is the case here, is a strong constraint which may be useful for constraining higher-loop results by their analytic properties (bootstrapping).  We remark that the quadrupole terms that couple four hard legs in the soft anomalous dimension matrix~\cite{Almelid:2015jia,Almelid:2017qju} have to obey this same constraint. Both cases involve four massless momentum vectors in a scattering process with (generically) more particles.  (In our case, one of the vectors is soft, but that does not matter for the definition of $z$ and $\bar{z}$ since they are rescaling invariant.)  In both cases the Minkowski kinematics can be smoothly continued around the real axis for $z>1$, and the physical result must be continuous during this process, implying a branch cut cancellation analogous to \eqn{discA1D2omzALT}, for example. On the other hand, the soft anomalous dimension quadrupole terms have a $z \leftrightarrow \bar{z}$ parity symmetry. (Parity is not a symmetry of the soft factor for emitting a positive-helicity gluon, because it maps it to negative-helicity emission.) After imposing parity in the Euclidean region, branch cut cancellation in Minkowski regions is likely to be automatic.
 
%=======================
\begin{figure}[ht!]
    \centering
    \begin{subfigure}[c]{0.1\textwidth}
      \raisebox{0pt}{\includegraphics[scale=0.4]
         {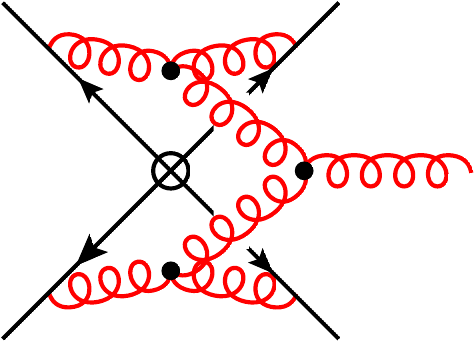}}
        \caption{}
        \label{fig:2a}
    \end{subfigure}
    \qquad \qquad
    \begin{subfigure}[c]{0.17\textwidth}
      \includegraphics[scale=0.4]{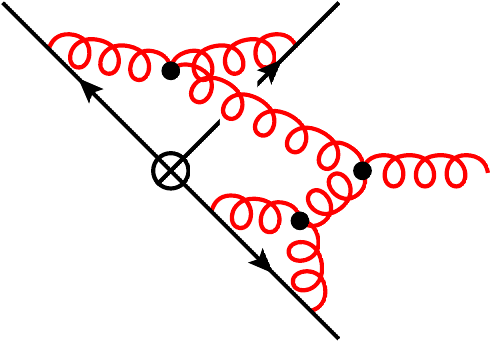}
        \caption{}
        \label{fig:2b}
    \end{subfigure}
   \qquad
    \begin{subfigure}[c]{0.1\textwidth}
      \includegraphics[scale=0.4]{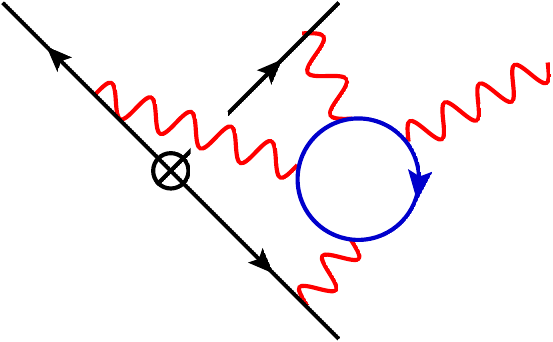}
        \caption{}
        \label{fig:2c}
    \end{subfigure}
    \caption{Representative diagrams for the (a) three-loop quadrupole,
      (b) three-loop tripole,
      and (c) three-loop Abelian soft factors.}
\label{fig:2}
\end{figure}
%=======================

Soft emission at higher loops should reveal additional structures.  For example, the three-loop soft emission factor will include a purely non-Abelian quadrupole term (see fig.~\ref{fig:2}(a)), whose finite part we expect to have a uniform transcendental weight 6, and nontrivial kinematic dependence on several cross ratios.  It will also contain a tripole term (fig.~\ref{fig:2}(b)) which will depend on the matter content, like the dipole term at two loops, but with kinematic dependence on $z_k^{ij}$ similar to the two-loop tripole term. We expect this term to have non-uniform weight, except in the case of $\mathcal{N}=4$ SYM.  Of course, the three-loop dipole term must have the same simple kinematic dependence as the one and two loop dipole terms.  Finally, in massless QED, the soft emission factor for unrenormalized amplitudes should get its first quantum correction at three loops, from a light-by-light scattering contribution shown in fig.~\ref{fig:2}(c). This Abelian contribution can have both tripole and dipole terms.  These terms cannot be associated with a tree-level soft emission factor because they have a different dependence on the charge of the hard fermion, versus the charge of the massless fermion in the loop.  Similar contributions in nonabelian gauge theory will involve the quartic Casimir operator.

In summary, we have presented the soft factor required for single
soft-gluon emission in two-loop gauge theory amplitudes. Applications
of our results include construction of infrared subtraction terms for
NNNLO jet cross sections (see also ref.~\cite{Catani:2019nqv} for the
tree-level triple soft emission contribution, and ref.~\cite{DelDuca:2019ggv} for the tree-level quadruple collinear splitting contribution),
the calculation of QCD corrections at NNNLO in the soft-gluon
approximation beyond leading-color,
the soft approximation of two-loop gauge theory amplitudes,
and the quantitative study of
potential
factorization violation.

\subsection*{Acknowledgments}
We thank Yu Jiao Zhu for checking part of the results in this paper.
We also thank Falko Dulat for analytic continuation assistance
and Claude Duhr, Andrew McLeod, and Giulia Zanderighi for useful discussions.
This work is supported by the Department of Energy under contract
DE-AC02-76SF00515. HXZ is supported by the National Natural Science
Foundation of China under contract No.~11975200. This research received
funding from the European Research Council (ERC) under the European
Union's Horizon 2020 research and innovation programme
(grant agreement No.~725110), ``Novel structures in scattering amplitudes''.
We thank Johannes Henn, Rourou Ma, Yongqun Xu and Yang Zhang for pointing out a mistake in a previous version of this paper.

\bibliographystyle{JHEP}
\phantomsection
%\bibliography{amp_refs}

\providecommand{\href}[2]{#2}\begingroup\raggedright\endgroup

\clearpage

\end{document}